%% file: main.tex
\documentclass[graybox, secnum]{svmult}
\pdfoutput=1
\usepackage[T1]{fontenc}
\usepackage[utf8]{inputenc}

\usepackage{mathptmx}       % selects Times Roman as basic font
\usepackage{helvet}         % selects Helvetica as sans-serif font
\usepackage{courier}        % selects Courier as typewriter font
\usepackage{type1cm}        % activate if the above 3 fonts are
\usepackage{makeidx}         % allows index generation
\usepackage{graphicx}        % standard LaTeX graphics tool
\usepackage{multicol}        % used for the two-column index
\usepackage[bottom]{footmisc}% places footnotes at page bottom
\usepackage{hyperref}        %for hyperlinks
\usepackage{soul}            % for high-lighting of text
\hypersetup{colorlinks=true,urlcolor=blue}
\usepackage[square,numbers]{natbib}

\usepackage{multirow}

\usepackage{hyperref}

\usepackage{mathtools}
\usepackage{amsmath,amssymb,amsfonts,latexsym,stmaryrd,physics,mathrsfs}

\allowdisplaybreaks[4]
\usepackage{marginnote}
\usepackage{bm}
\usepackage{float}
\usepackage{nicefrac}
\usepackage{microtype} 
\usepackage{booktabs}
\usepackage{verbatim}
\usepackage{setspace}

\usepackage{stfloats}
\usepackage{diagbox}
\usepackage{dsfont}

\usepackage{algorithm}
\usepackage{algorithmic}
\usepackage{listings}
\usepackage{textcomp}
\usepackage{url}
\usepackage{diagbox}
\usepackage{subfigure}

\usepackage[section]{placeins}

\input{def.tex}

\def\hbarn{{\mathchar'26\mkern-9muh}}

\graphicspath{{img/}}

\makeindex             % used for the subject index
\begin{document}
%\tableofcontents{}
\title*{Spinfoams and high performance computing}

\author{Pietro Dona, Muxin Han and Hongguang Liu}
\institute{Pietro Dona \at Center for Space, Time and the Quantum, 13288 Marseille, France,
\at Department of Physics and Astronomy, University of Western Ontario, London, ON N6A 5B7, Canada
\email{dona.pietro(AT)gmail.com}
\and Muxin Han \at Department of Physics, Florida Atlantic University, 777 Glades Road, Boca Raton, FL 33431-0991, USA, \at Department Physik, Institut f\"ur Quantengravitation, Theoretische Physik III, Friedrich-Alexander Universit\"at Erlangen-N\"urnberg, Staudtstr. 7/B2, 91058 Erlangen, Germany, \email{hanm(AT)fau.edu}
\and Hongguang Liu \at Department Physik, Institut f\"ur Quantengravitation, Theoretische Physik III, Friedrich-Alexander Universit\"at Erlangen-N\"urnberg, Staudtstr. 7/B2, 91058 Erlangen, Germany, \email{hongguang.liu(AT)gravity.fau.de}}
\maketitle
\abstract{Numerical methods are a powerful tool for doing calculations in spinfoam theory. We review the major frameworks available, their definition, and various applications. We start from \texttt{sl2cfoam-next}, the state-of-the-art library to efficiently compute EPRL spin foam amplitudes based on the booster decomposition. We also review two alternative approaches based on the integration representation of the spinfoam amplitude: Firstly, the numerical computations of the complex critical points discover the curved geometries from the spinfoam amplitude and provides important evidence of resolving the flatness problem in the spinfoam theory. Lastly, we review the numerical estimation of observable expectation values based on the Lefschetz thimble and Markov-Chain Monte Carlo method, with the EPRL spinfoam propagator as an example. }

\section*{Keywords} 
Loop Quantum Gravity, Spinfoam Theory, Numerical frameworks, High-performance computing, Booster decomposition, Lefschetz thimble.

%\tableofcontents

\section{Introduction}
\label{sec:intro}
\input{intro}

\section{Booster functions and SU(2) invariants: sl2cfoam}
\label{sec:booster}
\input{booster}

\section{Complex critical points and resolution of flatness problem}
\label{sec:mainflatness}
\input{flatness}

\section{Spinfoam Propagator and Lefschetz thimble}
\label{sec:Lefschetz}

\input{Lefschetz}

\section{Conclusions}
\label{sec:conclusions}
\input{conclusions}

\section{Acknowledgements}
The work of P.D. was made possible through the support of the  FQXi  Grant  FQXi-RFP-1818 and of the ID\# 61466 grant from the John Templeton Foundation, as part of the ``The Quantum Information Structure of Spacetime (QISS)'' Project (\href{qiss.fr}{qiss.fr}). P. D. also personally thanks Francesco Gozzini and Pietropaolo Frisoni for sharing some of the data used in Section~\ref{sec:booster}. M.H. receives support from the National Science Foundation through grants PHY-1912278 and PHY-2207763, and the sponsorship provided by the Alexander von Humboldt Foundation.
M.H. acknowledges IQG at FAU Erlangen-N\"urnberg, IGC at Penn State University, and Perimeter Institute for Theoretical Institute for the hospitality during his visits.   

\bibliography{refs.bib}
\bibliographystyle{jhep}

\end{document}

%% file: def.tex
\newcommand{\Z}{\mathbb{Z}}
\newcommand{\R}{\mathbb{R}}
\newcommand{\C}{\mathbb{C}}

\newcommand{\e}{\mathrm{e}}

\newcommand{\ccirc}{\kern0.2ex\vcenter{\hbox{$\scriptstyle\circ$}}\kern0.2ex}

\newcommand{\SLDC}{{\mathrm{SL}(2,\mathbb{C})}}

\newcommand{\Slc}{\mathrm{SL}(2,\mathbb{C})}

\def\be{\begin{eqnarray}}
\def\ee{\end{eqnarray}}

\newcommand{\ci}{\mathcal I}
\newcommand{\cj}{\mathcal J}

\newcommand{\cs}{\mathcal S}

\newcommand{\cz}{\mathcal Z}

\newcommand{\fa}{\mathfrak{a}}

\newcommand{\g}{\gamma}

\newcommand{\sig}{\sigma}

\renewcommand{\l}{\lambda}

\renewcommand{\O}{\Omega}
\renewcommand{\t}{\tau}

\newcommand{\rmd}{\mathrm d}

\newcommand{\lt}{\left}
\newcommand{\rt}{\right}

\newcommand{\sn}{\mathscr{N}}

\newcommand{\re}{\mathrm{Re}}

%% file: intro.tex
Spinfoam theory is a covariant formulation of Loop Quantum Gravity. It provides a background-independent, and Lorentzian quantum gravity path integral regularized on a fixed triangulation. Spinfoam assigns transition amplitudes to spin network states living on the triangulation's boundary. The state-of-the-art spinfoam model is the EPRL-FK model defined in \cite{Engle:2007wy,Freidel:2007py} (for a pedagogical introduction, see \cite{rovelli2014covariant, Perez:2012wv}). 

Computations in spin foam are very involved, and for a long time, the theory was relegated to a formal proposal, and very few explicit calculations existed. The majority of the results of the approach are obtained in the large spins regime, often identified as semiclassical. The theory shows a remarkable connection to a discrete version of General Relativity if the limit is taken simultaneously as the refinement. 

Recently, the field has undergone a numerical revolution. The interest of the community in numerical methods grew considerably while, at the same time, fast technological developments provided simpler access to High-Performance Computing facilities. 

Many different and complementary approaches were born, providing different tools suitable to solve different problems. We can finally do spinfoam computations and answer some of the theory open questions. In this work, we present a selection of these techniques.

The library \texttt{sl2cfoam} \cite{Dona:2018nev} was the first attempt to build a complete library to compute Lorentzian EPRL spinfoam transition amplitudes. It is coded in \texttt{C} and based on a divide-and-conquer paradigm. The amplitude is divided into easier-to-compute components and then reassembled. The library evolved into \texttt{sl2cfoam-next} \cite{Gozzini:2021kbt} and is now optimized for high-performance computing. With \texttt{sl2cfoam-next} an interactive \texttt{Julia} interface is provided.  This numerical framework is modular (can be used to compute any transition amplitude), scalable (can run on a laptop or a cluster), and user-friendly (a minimal amount of additional code is required). There are two main disadvantages of this approach. The number of resources required increases very rapidly but not exponentially. The numerical evaluation involves an unavoidable approximation (see Section~\ref{sec:truncation}), and we do not control it entirely.

The method of \emph{complex critical points} provides a complementary numerical approach for performing computations with the Lorentzian EPRL spinfoam model in the regime where the \texttt{sl2cfoam-next} code becomes computationally expensive is available. This approach is based on the integral expression of the spinfoam model. The main task is to compute oscillatory integrals representing the spinfoam amplitude numerically. This approach closely relates to the stationary phase approximation, one of the main tools for studying the quantum theory in the semiclassical regime. This numerical code focus on the complex critical points in the semiclassical regime of the spinfoam amplitude. The complex critical points are the key to recovering the curved geometries and the semiclassical gravity dynamics. The results produced by this code provide important pieces of evidence for resolving the confusion in the LQG community known as ``the flatness problem''  \cite{Engle:2020ffj,Hellmann:2012kz,Bonzom:2009hw,Perini:2012nd}. The flatness problem, which claims that the semiclassical geometries in the spinfoam model are all flat, results from mistakenly ignoring the contributions from complex critical points \cite{lowE,LowE1,lowE2}. This confusion is clarified by explicitly demonstrating the curved geometry from the spinfoam amplitude with the numerical method. 

From the computation of the spinfoam amplitude integrals beyond the leading order, we can derive the quantum corrections to the semiclassical limit of the theory\footnote{The quantum corrections derived from the spinfoam amplitude can also be analyzed numerically evaluating the next-to-leading order in the stationary phase approximation, see \cite{Han:2020fil}.}. The integrands in the spinfoam amplitude are highly oscillatory. Computing oscillatory integrals used to be numerically expensive, but the recent method based on the \emph{Lefschetz thimble} makes the computation much more efficient (see \cite{Alexandru:2020wrj} for a review). Given any critical point of the integral, the Lefschetz thimble is an integration cycle connected to the critical point, such that the integrand becomes non-oscillatory on the cycle and the original integral is a linear combination of the integrals on the Lefschetz thimbles. We developed a numerical code to compute the spinfoam correlation functions with the Lefschetz thimble and Monte-Carlo methods \cite{Han:2020npv}. Thanks to the non-oscillatory integrand on the Lefschetz thimble, computing the integral with the Monte-Carlo method becomes efficient \footnote{The Monte-Carlo method can also be used as a tool for finding critical points in the complexified integration space, see \cite{Huang:2022plb}.}. In addition to reproducing the correct semiclassical behavior of the spinfoam propagator, the numerical results efficiently compute the quantum corrections to the propagator from the spinfoam amplitude.

In addition to the numerical frameworks presented in this chapter, there exist other numerical codes based on the spinfoam models simplified compared to the EPRL model, such as the effective spinfoam model (see, e.g., \cite{Asante:2020iwm,Asante:2021zzh}), where %one obtains the complex critical points relating to curved geometries similar to the EPRL model
geometry and the connection with area-angles Regge calculus plays a central role, and the hypercuboid truncated model (see, e.g., \cite{Bahr:2017klw}), where the spinfoam renormalization is studied. We do not review them here since other chapters focus on these approaches. However, we would like to point out the interesting similarity between our result based on the EPRL model and the result from the effective spinfoam model. From the perspective of the semiclassical analysis, the effective spinfoam model also needs to apply the method of complex critical point and turns out to give qualitatively similar behavior to the result presented in Section \ref{sec:mainflatness}, particularly about the dependence of the amplitude on the curvature of the semiclassical geometry. It seems to suggest the close relationship between the effective spinfoam model and the EPRL model in the large-$j$ regime.

The architecture of this chapter is as follows: First, we discuss the library \texttt{sl2cfoam} in Section \ref{sec:booster}. Section \ref{sec:booster} also includes a concise review of the Lorentzian EPRL spinfoam amplitude and the booster function decomposition, which the library is based on. In Section \ref{sec:mainflatness}, we review the integral representation of the spinfoam amplitude, discuss the algorithm of computing the complex critical points, and demonstrate the curved geometries emergent from the spinfoam amplitude in a few numerical examples. In Section \ref{sec:Lefschetz}, we discuss the algorithm of computing the spinfoam oscillatory integrals on the Lefschetz thimble and the application on the spinfoam propagator.

%% file: booster.tex
In this section, we will review the main ingredients and a few applications of the booster decomposition of the EPRL amplitude and its numerical implementation in \texttt{sl2cfoam-next}. The reader interested in more details can find a recent and pedagogical introduction to this framework in \cite{Dona:2022dxs}. 

%
% ##################################################################################
%

\subsection{The EPRL transition amplitude}
This section briefly introduces the EPRL spin foam model and fixes our notation. For a comprehensive and more detailed description of the model, its derivation, physical motivation, and connection with LQG, we refer to other chapters of this book, reviews \cite{Perez2012}, or other monographs \cite{rovelli2014covariant}.

The EPRL spin foam theory attempts to quantize gravity with a path integral regularized on a triangulation of the spacetime manifold or, more precisely, its dual 2-complex $\Delta$. It provides dynamics to LQG, assigning a transition amplitude to the states in kinematical Hilbert space living at the boundary of the spinfoam.

The EPRL transition amplitude $A_\Delta$ is defined in terms of local quantities of the 2-complex $\Delta$ colored with LQG quantum numbers: each face with a spin $j_f$, and each edge with an intertwiner $i_e$. We have a face amplitude $A_f(j_f)$, an edge amplitude $A_e(i_e)$, and a vertex amplitude $A_v \left(j_f, \  i_e\right)$ each assigned to the corresponding component of the 2-complex. We also sum over all possible bulk quantum numbers.
\begin{equation}
\label{eq:partitionf}
A_{\Delta} = \sum_{j_f, i_e}  \prod_f A_f(j_f) \prod_e A_e(i_e) \prod_v A_v \left(j_f, \  i_e\right) \ ,
\end{equation}

The face and edge amplitudes are fixed, requiring the correct convolution property of the path integral at fixed boundary \cite{face} $A_f(j_f) = 2 j_f +1$ and $A_e(i_e)=2i_e+1$. The vertex amplitude is given by
\begin{equation}
    \label{eq:vertexamplitude}
    A_v \left(j_f, \  i_e\right)  = \sum_{m_s,m_t} 
    \int \prod_{e \in v} \D g_e 
    \delta(g_{e'}) \prod_{f \in v} D^{\gamma j_f, j_f}_{j_f m_{t}, j_f m_{s}} (g_{t}^{-1} g_{s})   
    \prod_{e \ni f} \left( \begin{array}{c}
        j_f\\
        m_p
    \end{array}\right)^{(i_e)} \ .
\end{equation}
The $D^{\gamma j_f, j_f}$ are the matrix elements unitary irreducible representations in the principal series of $\SLDC$ labeled by $\rho,k = \gamma j,j$ where $\gamma$ is the Immirzi parameter. 
The delta function $\delta(g_e')$ regularizes the amplitude removing a redundant integration as prescribed in \cite{finite}. %Each face of the vertex contributes with a $\gamma$-simple representation of $\SLDC$ of spin $j_f$ evaluated on t
The product of the holonomies $g_{t}^{-1} g_{s}$ represents the parallel transport from the source edge of the face $s$ to the target one $t$. We integrate over all of them. The magnetic indices $m_{s}$ and $m_{t}$
are contracted with an intertwiner tensor labeled by a quantum number $i_e$. % a Wigner $(4jm)$ symbol on each edge (we use a shorthand notation) labeled by the intertwiner quantum number $i_e$. 

This expression will be rewritten in a form tailored to each numerical technique in the following sections.  

%
% ##################################################################################
%

\subsection{Booster function decomposition of the amplitude}
Each face in the vertex amplitude \eqref{eq:vertexamplitude} is decorated with a spin $j_{f}$ and contributes to the amplitude with a $\gamma$-simple representation
\begin{equation}
D^{\gamma j_{f}, j_{f}}_{j_{f}m_{t}, j_{f}m_{s}} (g_t^{-1} g_s) = \sum_{l_{f}=j_f}^\infty \sum_{n_{f}=-l_f}^{l_f}
D^{\gamma j_{f}, j_{f}}_{j_{f}m_{t}, l_{f}n_{f}} (g_t^{-1} ) 
D^{\gamma j_{f}, j_{f}}_{l_{f}n_{f}, j_{f}m_{s}} (g_s) \ .
\end{equation}
We used the representation property to separate the contribution of the edge group elements. The sum over $l_{f}$ is bounded from below by $j_{f}$ and unbounded from above. General $SL(2,\C)$ unitary representations are infinite-dimensional. 

The vertex amplitude \eqref{eq:vertexamplitude} involves four integrations (one was removed for the regularization) over a six-dimensional non-compact group of highly oscillating functions. Performing brute-force integration is a complex and demanding task.

To simplify the integration, we parameterize each group element using the Cartan decomposition of $\SLDC$. The group element and integration measure decomposes as 
\begin{equation}
\label{eq:CartanSL2C}
    g = ue^{\frac{r}{2} \sigma_3} v^{-1} \ ,   \qquad \text{and} \qquad \D g =  \frac{1}{4\pi} \sinh^2 r \ \D r \ \D u\  \D v \ .
\end{equation}
where $u,v\in SU(2)$, and the rapidity $r\in [0,+\infty)$. This is analogous to choosing polar coordinates for an integral on the plane $\R^2$. The integral over two unbounded Cartesian coordinates is mapped in a bounded angular integration and an unbounded radial one. %The $\SLDC$ Haar measure decomposes accordingly
% \begin{equation}
% \label{eq:HaarSL2C}
%     \D g =  \frac{1}{4\pi} \sinh^2 r \ \D r \ \D u\  \D v \ ,
% \end{equation}
% where $\D u$ and $\D v$ are normalized Haar measures of $SU(2)$ and the factor $4\pi$ is taken as in \cite{Speziale:2016axj, ruhl}. 
The matrix elements in the unitary representations also decompose nicely with the parametrization \eqref{eq:CartanSL2C}.
\begin{equation}
    D^{\gamma j, j}_{ln , jm} (g) = \sum_{p=-j}^{j} D^{l}_{np} (u) d^{\gamma j, j}_{ljp} (r) D^{j}_{pm} (v^{-1}) \ , 
\end{equation}
where $d^{\gamma j, j}_{ljp} (r) \delta_{pq}= D^{\gamma j, j}_{lp , jq} (e^{\frac{r}{2} \sigma_3})$ is the reduced matrix element (see \cite{Speziale:2016axj,ruhl,Dona:2022dxs} for an expression in terms of hypergeometric functions).

Each edge group element contributes to the vertex amplitude with four matrix elements. The prototype of this contribution is 
\begin{equation}
    \int \D g_e  \prod_{f \ni a} D^{\gamma j_f ,j_f}_{l_f n_f j_f m_f} (g_e) = \int %\D u_e \D v_e  \frac{1}{4\pi} \sinh^2 r_e \ \D r_e  
    \D g_e \ \prod_{f \ni e} \sum_{p_f} D^{l_f}_{n_f p_f} (u_e) d^{\gamma j_f, j_f}_{l_fj_fp_f} (r_e) D^{j_f}_{p_f m_f} (v^{-1}_e)\ ,
\end{equation}
where $f$ labels the faces in the edge $e$. We perform the $SU(2)$ integrals in terms of $(4jm)$ Wigner symbols to obtain
\begin{equation}
    \int \D g_e  \prod_{f \ni e} D^{\gamma j_f ,j_f}_{l_f n_f j_f m_f} (g_e) =\sum_{i_e,k_e} (2 i_e+1)
    (2 k_e+1)
    \left(\begin{array}{c} l_f \\ p_f \end{array}\right)^{(k_e)} 
    B_4^\gamma \left( l_f, j_f ; i_e ,k_e\right)  
    \left(\begin{array}{c} j_f \\ p_f \end{array}\right)^{(i_e)} \ ,
\end{equation}
where we defined the booster function as the one dimensional integral
\begin{equation}
    \label{eq:boosterdef}
    B_4^\gamma \left( l_f, j_f ; i_e ,k_e\right) 
    =\sum_{ p_f} 
    \left(\begin{array}{c} l_f \\ p_f \end{array}\right)^{(k_e)}
    \left(\int_0^\infty \D r \frac{1}{4\pi}\sinh^2r \, \prod_{f \in e} d^{\gamma j_f,j_f}_{l_f j_f p_f}(r) \right)
    \left(\begin{array}{c} j_f \\ p_f \end{array}\right)^{(i_e)} 
    \ .
\end{equation}
Introduced in \cite{Speziale:2016axj}, the booster functions $B_4^\gamma$ are related to the Clebsh-Gordan coefficients of $\SLDC$ and can be evaluated in terms of complex gamma functions \cite{Anderson:1970ez, Kerimov:1978wf}, can be numerically computed with very high precision \cite{Dona:2018nev,Gozzini:2021kbt}, and possess an interesting geometrical interpretation in terms of boosted tetrahedra \cite{Dona:2020xzv}. The booster functions depend on the Immirzi parameter $\gamma$ and encode the quantum simplicity constraints for the EPRL model. 
The $(4jm)$ symbols involving the $l_f$ spins contract among themselves, forming a $\{15j\}$ symbols, a higher-order $SU(2)$ invariant. The other $(4jm)$ symbols contract with the corresponding $(4jm)$ symbols in the definition of the vertex amplitude \eqref{eq:vertexamplitude}.

In summary, we can rewrite the EPRL vertex amplitude \eqref{eq:vertexamplitude} into a linear combination of $\{15j\}$ symbols weighted by booster functions
\begin{align}
    \label{eq:amplitude-formula}
    A_{v} \left(j_f, \  i_e\right) = \sum_{l_f=j_f}^\infty \{ 15 j\} (j_f,l_f) \prod_{e=2}^5 B_4^\gamma \left( l_f, j_f ; i_e ,k_e\right) \ .
\end{align}

We reduced the problem of performing high-dimensional oscillatory integrals to computing a series of one-dimensional integrals, $SU(2)$ invariants, and summing all the elements. The expression \eqref{eq:amplitude-formula} for the vertex amplitude in the spin-intertwiner representation is particularly convenient for a numerical calculation since it consists of simpler and repeatable tasks. The library \texttt{sl2cfoam-next} provides a numerical implementation of the vertex amplitude computing booster functions and $\{15j\}$ symbols and summing them together in the most time and memory-efficient way.

%
% ##################################################################################
%

\subsection{SU(2) invariants}
The computation of SU(2) invariants can be very resource-intensive. Computing the $\{15j\}$ symbol as a contraction of $(4jm)$ symbols is inefficient in terms of memory usage and computational time. In \texttt{sl2cfoam-next} we compute the $\{ 15 j\}$ symbol of the first kind \cite{Yutsis} expressing it as a finite sum of $\{6j\}$ symbols. The optimized calculation of $SU(2)$ invariants is an important topic also in other scientific fields (spectroscopy, nuclear physics, or chemistry). We did not reinvent the wheel and use the very performant libraries \texttt{WIGXJPF} and \texttt{FASTWIGXJ} \cite{Johansson:2015cca} to compute the $\{6j\}$ symbols we need.

%
% ##################################################################################
%

\subsection{Booster functions}
One of the main advantages of the booster decomposition \eqref{eq:amplitude-formula} is reducing the group integrals to one-dimensional ones. Furthermore, it is possible to recast the reduced matrix elements of $\SLDC$ to finite sums of exponentials \cite{Dona:2018nev} and with a change of variable $r \mapsto e^{-r}$ map the unbounded integral to the unit interval $(0, 1]$. The drawback is dealing with highly oscillatory integrands. To evaluate the integral with high enough accuracy, we divide the interval $(0,1]$ into several subintervals depending on the spins in the booster function, the Immirzi parameter, and an optional accuracy parameter of the library. The subdivision is finer around the origin, where the integrand oscillates more. Finally, the integral over each subinterval is done using the Gauss-Kronrod quadrature method in double or quadruple precision.

%
% ##################################################################################
%
\subsection{The necessary approximation}
\label{sec:truncation}
A remnant of the non-compactness of the group in \eqref{eq:amplitude-formula} hides in the sums over the spins $l_f$. They are bounded from below by $j_f$ but not from above. Since the amplitude is finite \cite{finite}, the unbounded sums are convergent. To evaluate the amplitude numerically, we need to approximate the unbounded sums by truncating them. In \texttt{sl2cfoam-next} we choose to truncate all the summations homogeneously with the same parameter $\Delta l$, effectively replacing
\begin{equation}
    \sum_{l_f = j_f}^\infty \quad \longrightarrow \quad \sum_{l_f = j_f}^{j_f + \Delta l} \ .
\end{equation}

Unfortunately, we do not know how to estimate the error we are committing by truncating with a given $\Delta l$. Recently \cite{Frisoni:2021uwx, Dona:2022dxs}, we proposed to use convergence acceleration techniques to improve our approximation of the amplitude and reduce the dependence on the truncation parameter. Using Aitken's delta squared method, the vertex amplitude is well approximated by
\begin{equation}
    \label{eq:bounds-d4-1}
   A_{v} \approx \frac{A_{v}(\Delta l) A_{v}(\Delta l-2) - A_{v}^2(\Delta l-1)}{A_{v}(\Delta l) - 2A_{v}(\Delta l-1)+A_{v}(\Delta l-2)} \ .
 \end{equation}
The dependence of the result on other convergence techniques and the associated error is currently under investigation.

%
% ##################################################################################
%

\subsection{Assembling the amplitude}
To compute any EPRL transition amplitude, we perform many products (among invariants, booster functions, vertex amplitudes) and sums (spins, magnetic indices, intertwiners). For acceptable performances, we have to deal with them efficiently. The problem is similar to matrix multiplication in computer science, which is solved using specialized routines for contractions of multidimensional arrays and parallelization. 

To compute the vertex amplitude, we build a multidimensional array with the $\{15j\}$ symbols and another with the booster functions for varying spins $l_f$ and intertwiners. We calculate the vertex amplitude contracting the two tensors using basic linear algebra subprograms (BLAS) routines. The library is parallelized considerably using a hybrid OpenMP-MPI scheme. The calculation of booster tensor and vertex tensors for different $l_f$ is distributed on many MPI nodes, and every node parallelizes the necessary operations across its CPUs using OpenMP. The vertex amplitude is also stored in a multidimensional array, and the contraction of many vertices to form a transition amplitude can be parallelized.  For increased performance, tensor contractions can be parallelized using GPU cores with the \texttt{julia} module \texttt{SL2CfoamGPU}.

%
% ##################################################################################
%

\subsection{Applications}
We used \texttt{sl2cfoam} and \texttt{sl2cfoam-next} to compute numerous transition amplitudes. In this section, we review some of the main results. 

\subsubsection{Quickstart}
\begin{figure}[H]
    \centering
    \includegraphics[width=\linewidth]{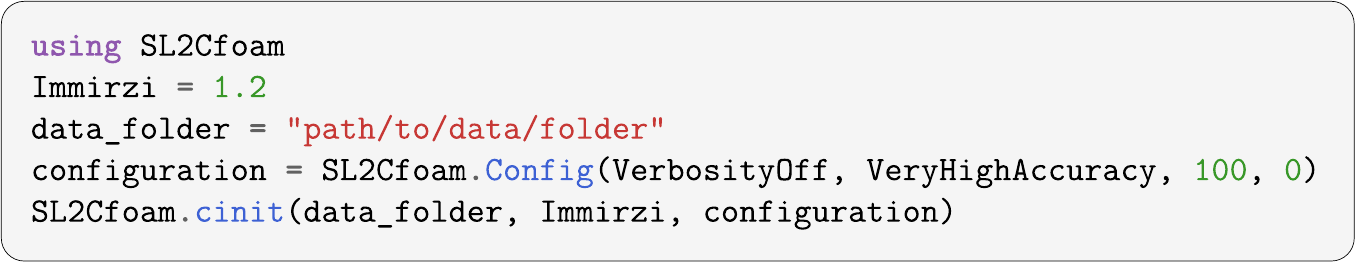}
    \caption{Initialization of \texttt{sl2cfoam-next}}
    \label{code-init}
\end{figure}

We start with a showcase of how user-friendly the library is. We need a few lines of code to compute one vertex amplitude using the \texttt{julia} interface. In Figure~\ref{code-init} we show how to import and initialize the library.

% \begin{jllisting}[caption={Initialization of \texttt{sl2cfoam-next}},label={code-init},linewidth=\textwidth]
%     using SL2Cfoam
%     Immirzi = 1.2
%     data_folder = "path/to/data/folder"
%     configuration = SL2Cfoam.Config(VerbosityOff, VeryHighAccuracy, 100, 0)
%     SL2Cfoam.cinit(data_folder, Immirzi, configuration) 
% \end{jllisting}

\begin{figure}[H]
    \centering
    \includegraphics[width=\linewidth]{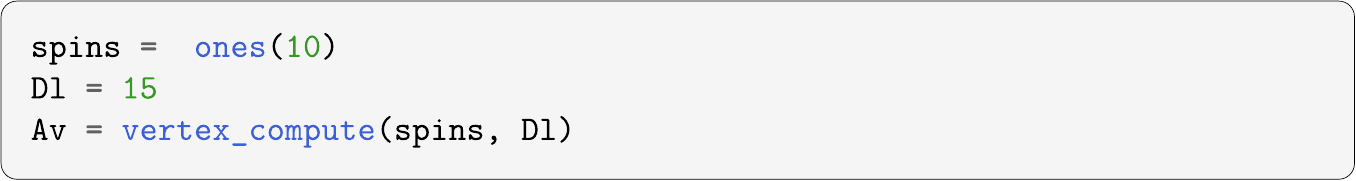}
    \caption{Computation of a vertex amplitude with \texttt{sl2cfoam-next}}
    \label{code1}
\end{figure}

We specify the value of the Immirzi parameter (set to $1.2$ in this example) and a folder where the library stores and retrieves the \texttt{FASTWIGXJ} tables of invariants and the computed amplitudes. Saving the value of known amplitudes and avoiding recalculations saves a lot of computational time in the long run in exchange of disk space. In Listing~\ref{code1}, we compute the vertex amplitude.

% \begin{jllisting}[caption={Computation of a vertex amplitude with \texttt{sl2cfoam-next}},label={code1}]
%     spins =  ones(10)
%     Dl = 15
%     Av = vertex_compute(spins, Dl)
% \end{jllisting}

We set the ten boundary spins (in this example, all equal to 1) and the value of the truncation parameter (in this example $\Delta l=15$). Finally, we compute the vertex amplitude and store it in an array with five indices, one per intertwiner. 

\subsubsection{Single vertex asymptotic}
The most celebrated result of the EPRL spinfoam model is its connection with the Regge action in the large spin limit with coherent boundary data \cite{Barrett:2009mw,HZ, Dona:2019dkf}.
\begin{equation}
    \label{eq:coherentvertex}
    A_v(j_f,\vec{n}_f) = \sum_{i_e} A_v(j_f,i_e) \prod_e c_{i_e} (\vec{n}_f) \ .
\end{equation}
The vertex amplitude \eqref{eq:amplitude-formula} is contracted with Livine-Speziale coherent intertwiners $c_{i_e}(\vec{n}_f)$ \cite{LS} representing the spacelike boundary of a Lorentzian 4-simplex. When the spins are homogeneously large, i.e., under a rescaling $j_f \to \lambda j_f$ with $\lambda \gg 1$, the amplitude \eqref{eq:coherentvertex} takes the asymptotic form
\begin{equation}
    \label{eq:asymptotic}
    A_v(\lambda j_{f}, \vec{n}_{f}) = \frac{1}{\lambda^{12}} (N_1 e^{i \lambda S_R} + N_2 e^{-i\lambda S_R}) + O(\lambda^{-13}) \ ,
\end{equation}
where $S_R$ is the Regge action of the $4$-simplex
\begin{equation}
    S_R = \sum_{f} \gamma \theta_{f}(\vec{n}_{f})  j_{f} \ ,
\end{equation}
with $\gamma$ the Immirzi parameter, $\theta_{f}$ are the dihedral angles of the $4$-simplex. The coefficients $N_s$ can be computed exactly, but a simple closed formula in terms of the 4-simplex geometry does not exists \cite{Dona:2019dkf,Han:2020fil}. 

The path towards the numerical verification of the formula \eqref{eq:asymptotic} started in \cite{Dona:2019dkf} with \texttt{sl2cfoam} and was completed in \cite{Gozzini:2021kbt} with \texttt{sl2cfoam-next}. This was the perfect testing ground for the library since it is one of the few analytic results in the theory.

We use boundary data representing a very symmetric isosceles Lorentzian 4-simplex (see \cite{Dona:2019dkf} for a detailed description of the boundary data). We focus on the Immirzi parameter $\gamma=2.0$, scales up to $\lambda=30$ with a truncation $\Delta l = 8$. 

\begin{figure}[H]
    \centering
    \includegraphics[width=0.8\linewidth]{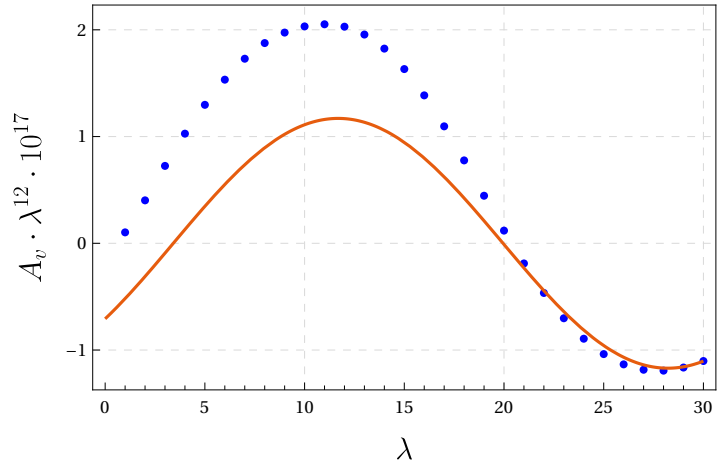}
    \caption{Plot of the rescaled coherent vertex amplitude \eqref{eq:coherentvertex} as function of the scale $\lambda$. The numerical evaluation uses $\gamma=2$ and $\Delta l =8$ and boundary data corresponding to a Lorentzian isosceles 4-simplex. The asymptotic expression \eqref{eq:asymptotic} with $N_1=\bar{N_2}= 5.85\cdot 10^{-18} \cdot e^{i 2.22}$ and $S_R = -0.19$ is plotted as a red continuous line.  The data was kindly shared by the authors of \cite{Gozzini:2021kbt}.}
    \label{fig:vertexasymptotic}
\end{figure}

We summarize the results in Figure~\ref{fig:vertexasymptotic}. The numerical evaluation and the asymptotic formula are in excellent qualitative and quantitative agreement, even at relatively small scales. 

This analysis consecrated \texttt{sl2cfoam-next} as a priceless tool to study the EPRL spinfoam model. However, it also highlights one of the limitations of the framework. The larger the scale of the boundary spins $\lambda$, the larger truncation $\Delta l$ is needed to approximate the amplitude. This agreement was possible only thanks to the optimizations introduced in \texttt{sl2cfoam-next}. A similar calculation performed in \cite{Dona:2019dkf} using \texttt{sl2cfoam} was limited to $\Delta l =1$, and only the order of magnitude of the numerical data and the asymptotic formula agreed. 

For other asymptotic behaviors of \eqref{eq:amplitude-formula} a similar analysis was already performed in \cite{Dona:2019dkf} using \texttt{sl2cfoam} obtaining already a good qualitative and quantitative match.

\subsubsection{Many vertices and the flatness problem}
\label{Many vertices and the flatness problem}

The regime of the EPRL spinfoam theory in which we can recover (discrete) general relatively is still under study. The presence of the Regge action in the large spin limit of a single vertex amplitude is not enough. We must explore extended triangulations to find if we can also extract the Regge equation of motion. The community agrees that a double scaling limit of low energies and refined triangulation (corresponding to large spins and many vertices) is the best candidate \cite{LowE1,lowE2,Han:2017xwo,Asante:2020iwm,Engle:2021xfs}.

Numerical calculations played an essential role in confirming the so-called flatness problem of the EPRL model \cite{CFsemiclassical,Bonzom:2009hw,frankflat,lowE2,Engle:2020ffj}. 
For a path integral formulation of a quantum theory, the amplitude is exponentially suppressed if and only if the boundary data is inconsistent with the classical equation of motions.
The flatness problem claims that, at fixed triangulation, the amplitude is dominated in the large spin limit by flat geometries. It was interpreted as a signal that the EPRL model apparently cannot recover non-flat solutions of Einstein's equations. The problem disappears if a refinement of the triangulation is taken into account at the same time as the large spin limit (see Section~\ref{sec:flatness} for a complementary discussion).   

\begin{figure}[H]
   \centering
   \raisebox{-0.5\height}{\includegraphics[width=0.8\linewidth]{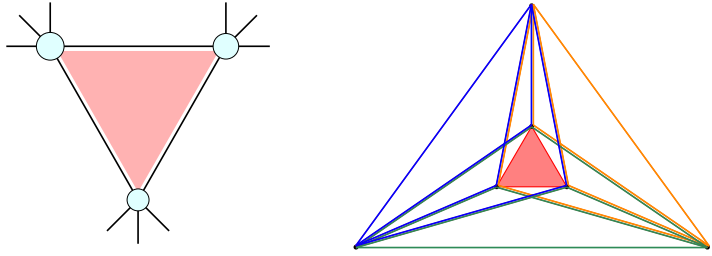}}
   \caption{
   \label{fig:D3pic} Left panel: The spin foam diagram associated to the $\Delta_3$ triangulation. We highlighted in red the face dual to the bulk triangle. Right panel: The $\Delta_3$ triangulation. We highlighted in red the bulk triangle and colored the three 4-simplices for a simpler visualization.}
\end{figure}

It is possible to verify numerically that the Lorentzian EPRL transition amplitude with coherent boundary data compatible only with a curved geometry exponentially decreases when the boundary spins are rescaled homogeneously. We tested it with the simplest triangulation $\Delta_3$ consisting of three 4-simplices sharing a tetrahedron and all sharing one triangle dual to the only bulk face (see Figure~\ref{fig:D3pic}). 

The amplitude associated with the $\Delta_3$ triangulation is 
\begin{equation}
\begin{split}
A_{\Delta_3}(j_f,\vec{n}_f)  = \sum_{i_e} \sum_{i_h} \sum_{j_{h}}(2 j_h +1) & A_{v_1}(j_f,i_e; j_h , i_{h_1}, i_{h_2})A_{v_2}(j_f,i_e; j_h , i_{h_2}, i_{h_3}) \times \\
& A_{v_3}(j_f,i_e; j_h , i_{h_3}, i_{h_1}) \prod_e c_{i_e} (\vec{n}_f) \ ,
\end{split}
\end{equation}
where we denoted with $h$ the face dual to the bulk triangle (highlighted in red in Figure~\ref{fig:D3pic}), with $j_h$ the associated spin, with $h_1$, $h_2$, $h_3$ the three bulk edges, with $i_{h_1}$, $ i_{h_2}$, $ i_{h_3}$ the associated intertwiners, and with $c_{i_e}$ the boundary coherent states representing a classical curved geometry. 

For simplicity, we considered highly symmetric boundary data with all the spins equal $j_f = j$. The rate of exponential decrease at large spins is related to the curvature of the geometry: the more curved the geometry is, the fastest the amplitude decreases. We constructed boundary data corresponding to a euclidean curved geometry with the largest possible (discrete) curvature compatible with the symmetry (see \cite{Dona:2020tvv} for a detailed description of the boundary data).

\begin{figure}[H]
   \centering
   \raisebox{-0.5\height}{\includegraphics[width=0.8\linewidth]{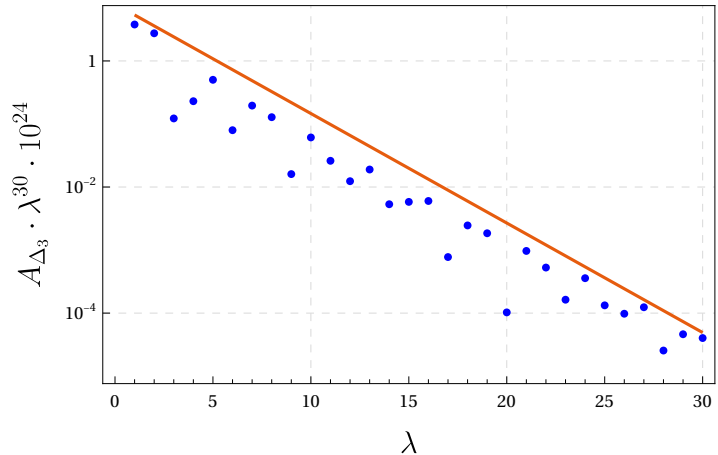}}
   \caption{
   \label{fig:delta3} Log-linear plot of the absolute value of the $A_{\Delta_3}$ EPRL amplitude for different boundary spins scale $\lambda$. The calculation is done with Immirzi parameter $\gamma=2$ and small truncation parameter $\Delta l = 2$. We plot the function $8 e^{-4 \lambda/10}$ with a continuous blue line. This function is just a visual guide highlighting the amplitude's exponential decrease. It is not the result of a fit, and we do not have enough data points to disentangle the exponential behavior from the oscillations. The data was kindly shared by the authors of \cite{Gozzini:2021kbt}. }
\end{figure}

In Figure~\ref{fig:delta3}, we plot the EPRL transition amplitude for the $\Delta_3$ triangulation with Immirzi parameter $\gamma=2$, truncation parameter $\Delta l = 2$, and rescaling the boundary spins homogeneously with scale $\lambda$ from $1$ to $30$. The exponential decrease in the amplitude is apparent. Thus we confirm that just taking large spins is not a good semiclassical limit of the theory.

\subsubsection{Radiative corrections}
Spinfoams are ultraviolet finite. However, in the case of a vanishing cosmological constant, they are affected by infrared divergences. Divergences arise due to unbounded summation over the spin labels on bulk faces. They require some renormalization procedure, and in general, their study and understanding are essential in defining the continuum limit.

The divergences of the Lorentzian EPRL model have been studied analytically using asymptotic techniques in \cite{Riello2013}, with a hybrid numerical and analytical analysis in \cite{Dona:2018pxq}, and very recently numerically using \texttt{sl2cfoam-next} with a massive computational effort in \cite{Frisoni:2021uwx}.

We focus on the computation of the radiative corrections to the EPRL spinfoam edge. The amplitude comprises two vertices glued along four edges, with six internal and four boundary faces. We sum over the spins of the internal faces $j_{h}$, and we fix the spins to the boundary faces $j_{f}$. Similarly, we sum over the bulk intertwiners associated with the internal edges $i_{h_e}$, and we fix the boundary ones  $i_{e}$. We regularize the two vertex amplitudes removing the same integration over the boundary edge. The result is two identical vertex amplitudes, and the radiative correction amplitude is given by
\begin{equation}
\label{eq:radiativecorrection}
A_{rc} = \sum_{j_{h}} \left(\prod_{f\in h} (2 j_f +1 ) \right)\sum_{i_{h_e}} \left(\prod_{h_e} (2 i_{h_e} +1 ) \right)  A_v^2(j_{h}, j_{f},i_{h_e}, i_{e}) \ .
\end{equation}
The sums over $j_{h}$ are unbounded and can cause the amplitude to diverge. We introduce a homogeneous cutoff $K$ and study the value of the amplitude as a function of $K$. The nature of this cutoff is utterly different from the truncation $\Delta l$, introduced as a tool to approximate the convergent sums over the virtual spins $l_f$ and the vertex amplitudes $A_v$. 

\begin{figure}[H]
    \centering
    \raisebox{-0.5\height}{\includegraphics[width=0.8\linewidth]{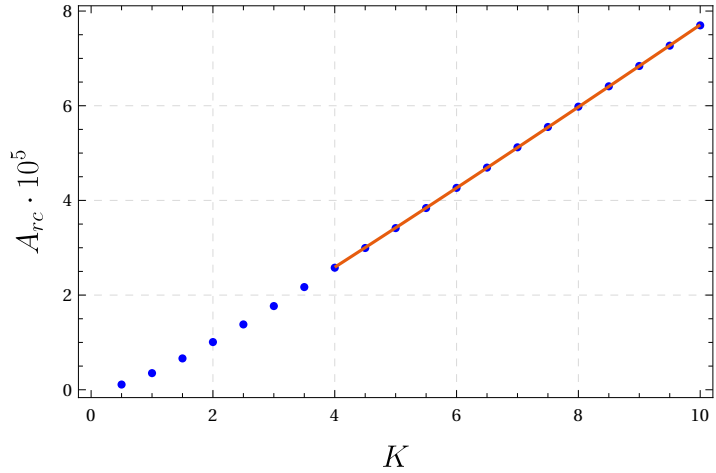}}
    \caption{Plot of the radiative correction amplitude $A_{rc} \times 10^5$ as a function of the cutoff $K$ in halfinteger steps.  The numerical evaluation uses $\gamma=0.1$ and $\Delta l =20$. The data was kindly shared by the authors of \cite{Frisoni:2021uwx}. }
    \label{fig:bubble}
\end{figure}

We numerically evaluate the amplitude \eqref{eq:radiativecorrection} for values of the cutoff $K$ from $\frac{1}{2}$ to $10$ at half-integers steps. We fix the boundary spins to $j_f = \frac{1}{2}$ and the boundary intertwiners to $i_e=0$. For each $K$ we fix $\Delta l = 20$, and we approximate the amplitude using the convergence acceleration formula \eqref{eq:bounds-d4-1}. In this calculation, we use $\gamma = 0.1$. We summarize the results in Figure~\ref{fig:bubble}.

We fit the amplitude with a function $a + b K^c$ and find $c=1.06$. We confidently claim that also the amplitude is scaling linearly. For completeness we report also the values of $a=-5.31 \cdot 10^{-6}$ and $b=7.18 \cdot 10^{-6}$.

Recently, in \cite{Dona:2022vyh}, other potentially divergent contributions of the radiative corrections to the EPRL propagator have been studied. All the contributions of diagrams with four or two bulk faces are indeed convergent. The results with different boundary data and Immirzi parameters are incredibly stable.  

\subsubsection{Other applications and future developments}
The numerical framework \texttt{sl2cfoam-next} is mainly used in phenomenological applications: to study the early universe, quantum tunneling of a black hole to a white hole, and explore the continuum limit. 

Some preliminary results are available \cite{Gozzini:2019nbo, Frisoni:2022urv}. In this section, we would like to mention some innovative techniques used to overcome the issue of \texttt{sl2cfoam-next} of being very resource-intensive on large triangulations. 

For example, to calculate the expectation value of a geometrical operator $O(j_f,i_e)$ at fixed boundary spins $j_f$, it is necessary to compute the spin-foam amplitude for every value of the boundary intertwiners $i_e$. With $20$ boundary tetrahedra, all boundary spins equal $j$, and the computation of one expectation value requires the evaluation of $(2j+1)^{20}$ amplitudes. Even in the small spins regime $j=2$, and with the most optimistic estimate of $1\mu s$ of CPU time per amplitude, the computation of one expectation value would take $5^{20} \mu s \approx 3$ years. 

The proposed solution to this problem uses statistical methods like Markov chain Monte Carlo techniques to evaluate the many sums and efficiently sample the space of possible amplitudes.

A similar issue also arises when we want to compute a spin foam transition amplitude with many bulk faces. The amplitude calculation must be repeated for many values of the bulk variables, and the CPU time needed piles up rapidly.

Another actively developed direction is the reduction of the dependence on the numerical result from the truncation parameters. Other convergence acceleration techniques and a systematic way to estimate the error in the amplitude are being studied.

%% file: flatness.tex
The semiclassical consistency is an important requirement in quantum physics, as any satisfactory quantum theory must reproduce the corresponding classical theory in the approximation of small $\hbarn$. In particular, the role of semiclassical analysis is crucial in quantum gravity, as it is the zero-th order test of the consistency with General Relativity. Moreover the semiclassical expansion provides a method of computing quantum effect perturbatively and connects the non-pertubative formulation of LQG to the perturbative regime. In spinfoam models, the semiclassical regime relates to the large-$j$ regime. As mentioned above, the numerical framework \texttt{sl2cfoam-next} is very resource-intensive for computing the spinfoam amplitude in the large-$j$ regime and for large triangulations. Therefore, we look for the alternative strategies toward the semiclassical analysis of spinfoam models. In this section, we introduce the numerical method based on the stationary phase approximation of the spinfoam amplitude, and in particular, we discussion the method of complex critical point, which is crucial for obtain curved geometries from the large-$j$ spinfoam amplitude. The discussion of this section is mostly based on the recent result in \cite{Han:2021kll}. Some earlier literature on the stationary phase analysis of spinfoam amplitude are  \cite{Conrady:2008mk,Barrett:2009mw,Han:2011re,Han:2013gna,Kaminski:2017eew,Liu:2018gfc,Simao:2021qno,Dona:2020yao}.

\subsection{Spinfoam amplitude and complex critical points}

\subsubsection{Integral representation of spinfoam amplitude}\label{sec_integral}

The 4-dimensional triangulation $\Delta$ are made by 4-simplices $v$, tetrahedra $e$, triangles $f$, line segments, and points. Here we denote the internal triangle by $h$ and the boundary triangle by $b$ ($f$ is either $h$ or $b$), and assign the SU(2) spins $j_h,j_b\in\mathbb{N}_0/2$ to $h,b$. The Lorentzian EPRL spinfoam amplitude on $\Delta$ sums over internal spins $\{j_h\}$ and has the following integral expression in analogy with path integral:
\be
A(\Delta)&=&\sum_{\{j_{h}\}}^{j^{\rm max}}\prod_h \bm{d}_{j_h}\int [\rmd g\rmd\mathbf{z}]\, e^{S%_{b[\vec{j}_b,\{\xi_b\}]}
\left(j_{h}, g_{v e}, \mathbf{z}_{vf};j_b,\xi_{eb}\right)}. \label{amplitude}
\ee
$\bm{d}_{ j_h}=2j_h+1$ is the dimension of SU(2) irreducible representation. The boundary states of $A(\Delta)$ are the SU(2) coherent states $|j_b,\xi_{eb}\rangle$, where $j_b,\xi_{eb}$ determines the area and the 3-normal of $b$ in the boundary tetrahedra $e$. The summed/integrated variables are $g_{ve}\in\Slc$, $\mathbf{z}_{vf}\in\mathbb{CP}^1$, and $j_h$, while the boundary $j_b,\xi_{eb}$ are not summed/integrated. The integration measure is $[\rmd g\rmd \mathbf{z}]=\prod_{(v, e)} \mathrm{d} g_{v e} \prod_{(v,f)} \mathrm{d}\O_{\mathbf{z}_{v f}}$, where $\rmd g_{ve}$ is the $\Slc$ Haar measure, and $\mathrm{d}\O_{\mathbf{z}_{v f}}$ is a scaling invariant measure on $\mathbb{CP}^1$. We truncate the sum over internal $j_h\in\mathbb{N}_0/2$ at $j^{\rm max}$. $j^{\rm max}$ is determined by boundary spins $j_b$ via the triangle inequality for some internal triangles, otherwise $j^{\rm max}$ is an IR cut-off. The spinfoam action $S$ is complex and linear to the spins $j_h,j_b$. Here we emply the action $S$ derived in \cite{Han:2013gna}, so that there is no internal $\xi_{eh}$ appearing in the integral of $A(\Delta)$ (this version of integral expression of $A(\Delta)$ turns out to avoid some degeneracy of the Hessian). We skip the detailed expression of $S$ but refer the reader to \cite{Han:2013gna} (see also \cite{Han:2021kll}).
%\be
%S%_{b[\vec{j}_b,\{\xi_b\}]}\left[j_{f}, g_{v e}, z_{v f}\right]
%&=&\sum_{e'}j_hF_{(e',h)}+\sum_{(e,b)}j_bF^{in/out}_{(e,b)}+\sum_{(e',b)}j_bF^{in/out}_{(e',b)},\label{SjFjF} 
%\\
%F_{(e,b)}^{out}&=&2 \ln\dfrac{\left\langle Z_{v e b},\xi_{e b}\right\rangle}{\left\| Z_{v e b}\right\|}+i\g \ln \left\| Z_{v e b}\right\|^2,\\
%F_{(e,b)}^{in}&=& 2 \ln \dfrac{\left\langle\xi_{e b}, Z_{v' e b}\right\rangle}{\left\| Z_{v' e b}\right\|}-i\g \ln \left\| Z_{v' e b}\right\|^2,\\
%F_{(e',f)}&=&2 \ln \dfrac{\left\langle Z_{v e' f}, Z_{v^{\prime} e' f}\right\rangle}{\left\| Z_{v e' f}\right\|\left\| Z_{v^{\prime} e' f}\right\|} + i\g \ln\frac{\left\| Z_{v e' f}\right\|^2}{\left\| Z_{v^{\prime} e' f}\right\|^2}.
%\ee
%$Z_{vef}=g^\dagger_{ve}\mathbf{z}_{vf}$ and $f=h$ or $b$. $e$ and $e'$ are the boundary and internal tetrahedra respectively. We denote the dual complex by $\Delta^*$. The orientation of the face $f^*$ dual to $f$ induces $\partial f^*$'s orientation that is outgoing from the vertex dual to $v$ and incoming to another vertex dual to $v'$. The logarithms are fixed to be the principle value. 

We apply the Poisson summation formula to change the sum over $j_h$ to the integral, as a preparation for the stationary phase analysis. Indeed, we firstly replace each $\bm{d}_{j_h}$ by a smooth compact support function $\t_{[-\epsilon,j^{\rm max}+\epsilon]}(j_h)$ satisfying
\be
&&\t_{[-\epsilon,j^{\rm max}+\epsilon]}( j_h)=\bm{d}_{j_h},\quad j_h\in[0,j^{\rm max}]\quad \text{and}\quad \t_{[-\epsilon,j^{\rm max}+\epsilon]}(j_h)=0,\nonumber\\
&&j_h\not\in[-\epsilon,j^{\rm max}+\epsilon],
\ee
for any $0<\epsilon<1/2$. This replacement does not change the value of the amplitude $A(\Delta)$, but now the summand of $\sum_{j_h}$ is smooth and compact support in $j_h$. We apply the Poisson summation formula 
\[
\sum_{n\in\Z} f(n)=\sum_{k \in \mathbb{Z}} \int_\R \mathrm{d} n f(n) \,{e}^{2\pi i k n},
\]
to the spinfoam amplitude. The discrete sum over $j_h$ in $A(\Delta)$ becomes the integral,
\be
A(\Delta)&=&\sum_{\{k_h\in\mathbb{Z}\}} \int_\R%\limits_{-\epsilon}^{j^{\rm max}+\epsilon} 
\prod_h\mathrm{d} j_{h}\prod_h 2\t_{[-\epsilon,j^{\rm max}+\epsilon]}(j_h)\int [\rmd g\rmd \mathbf{z}]\, e^{S^{(k)}},\nonumber\\
&&S^{(k)}=S+4\pi i \sum_h j_h k_h.\label{integralFormAmp1}
\ee

By the area spectrum of LQG $\fa_f=8\pi\g G\hbarn\sqrt{j_f(j_f+1)}$, the classical area $\fa_f$ and small $\hbarn$ imply the large spin $j_f\gg1$. This motivates to understand the large-$j$ regime as the semiclassical regime of the spinfoam model. To probe the semiclassical regime, we scale uniformly the boundary spins $j_b\to \l j_b$ with any $\l\gg1$, and make the change of variables for the internal spins $j_h\to\l j_h$, at the same time, we also scale $j^{\rm max}$ by $j^{\rm max}\to \l j^{\rm max}$. The resulting $A(\Delta)$ is given by  
\be
A(\Delta)&=&\sum_{\{k_h\in\mathbb{Z}\}}\int_\R%\limits_{-\epsilon/\l}^{j^{\rm max}+\epsilon/\l} 
\prod_h\mathrm{d} j_{h}\prod_h 2 \l\,\t_{[-\epsilon,\l j^{\rm max}+\epsilon]}(\l j_h)\int [\rmd g\rmd \mathbf{z}]\, e^{\l S^{(k)}},\nonumber\\
&&S^{(k)}=S+4\pi i \sum_h j_h k_h,\label{integralFormAmp22}
\ee
where $j_h$ is real and continuous. Each term in $A(\Delta)$ at fixed $\{k_h\}$ is cast into a standard form ready for the stationary phase approximation, in order to explore its semiclassical behavior as $\l\gg1$.

\begin{figure}[h]	
	\centering
	\includegraphics[scale=0.25]{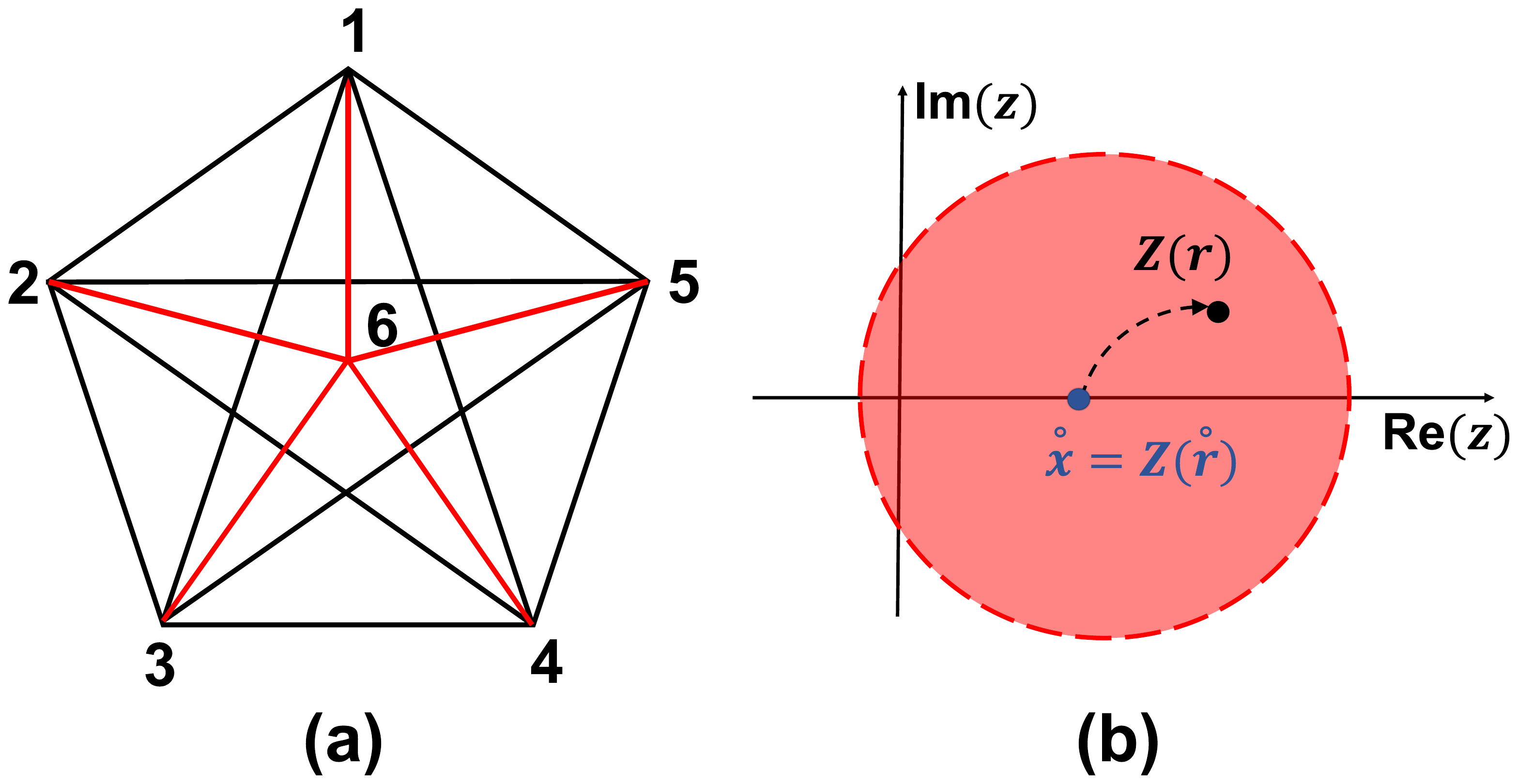}
	\caption{
	%(a) The $\Delta_3$ triangulation (the center panel) made by gluing three 4-simplices (in blue, red, and purple). The internal triangle $(135)$ is highlighted in red. 
	(a) The triangulation $\sig_{\text{1-5}}$ made by the 1-5 Pachner move dividing a 4-simplex into five 4-simplices. $\sig_{\text{1-5}}$ has 10 internal triangles and 5 internal segments $I=1,\cdots,5$ (red). (b) The real and complex critical points $\mathring{x}$ and $Z(r)$. $\cs(r,z)$ is analytic extended from the real axis to the complex neighborhood illustrated by the red disk.}
	\label{pic}
\end{figure}

\subsubsection{The flatness problem}

By the stationary phase approximation, each integral in \eqref{integralFormAmp22} with $\l\gg1$ receives the dominant contributions from solutions of the critical equations 
\be
\re(S)&=&\partial_{g_{ve}}S=\partial_{\mathbf{z}_{vf}}S=0,\label{eom1}\\
\partial_{j_h}S&=&4\pi i k_h, \qquad k_h\in\Z.\label{eom2}
\ee
We view the integration domain as a real manifold, and the solution inside the integration domain, denoted by $\{\mathring{j}_h,\mathring{g}_{ve},\mathring{\bf z}_{vf}\}$, is refered to as the \textit{real critical point}. 

Every real solution satisfying the part \eqref{eom1} and a nondegeneracy condition endows a Regge geometry to $\Delta$ \cite{Han:2013gna,Han:2011re,Barrett:2009mw,Conrady:2008mk}. However, further imposing \eqref{eom2} to these Regge geometries gives the accidental flatness constraint to every deficit angle $\delta_h$ hinged by the internal triangle $h$ \cite{Bonzom:2009hw,Han:2013hna}
\be
\g \delta_h= 4\pi k_h, \qquad k_h\in\Z. \label{flat}
\ee
The Barbero-Immirzi parameter $\g\neq 0$ is finite. When $k_h=0$, $\delta_h$ at every internal triangle is zero, so the Regge geometry endowed by the real critical point is flat. If the dominant contribution to $A(\Delta)$ with $\l\gg1$ only came from real critical points, Eq.\eqref{flat} would imply that only the flat geometry and geometries with $\g\delta_h=\pm 4\pi\mathbb{Z}_+$ could contribute dominantly to $A(\Delta)$, whereas the contributions from generic curved geometries were suppressed. Then the semiclassical behavior of $A(\Delta)$ would fail to be consistent with GR. By the argument in \cite{LowE1,lowE} and confirmed recently by \cite{Han:2021kll}, the resolution of this flatness problem requires to analytically extend the equations \eqref{eom1} and \eqref{eom2} to the complexified variables and solve for the \emph{complex critical points}, which turn out to encode curved geometries.

Before coming to the details about complex critical point, we would like to mention that a generic $\{\mathring{j}_h,\mathring{g}_{ve},\mathring{\bf z}_{vf}\}$ can endow discontinuous 4d orientation to $\Delta$, i.e., the orientation flips between 4-simplices. Then generally \eqref{flat} may become $\g \sum_{v\in h}s_v\Theta_{h}(v)= 4\pi k_h$ where $s_v=\pm1$ labels two possible orientations at each $4$-simplex $v$. $\Theta_{h}(v)$ is the dihedral angle hinged by $h$ in $v$.

\subsubsection{Complex critical points}\label{complex_critical_point}

The argument toward the flatness problem assumes all the dominant contribution to the spinfoam amplitude comes from the real critical point. However, as we will show, the large-$\l$ spinfoam amplitude does receive dominant contributions from the complex critical point away from the real integration domain. The complex critical points encode the curved geometries missing in the above argument. Demonstrating this property requires a more refined stationary phase analysis: We come back to the amplitude \eqref{integralFormAmp22} and separate $M$ internal areas $j_{h_o}$ ($h_o=1,\cdots,M$) from other $j_{\bar{h}}$ ($\bar{h}=1,\cdots,F-M$). $F$ is the total number of internal triangles in $\Delta$. $M$ equals to number of internal segments $I$. The areas $\{j_{h_o}\}$ are suitably chosen such that we can change variables from the areas $\{j_{h_o}\}_{h_o=1}^M$ to the internal segment-lengths $\{l_I\}_{I=1}^M$ (by inverting Heron's formula \footnote{We relate the chosen $M$ areas $\{j_{h_o}\}$ to $M$ segment-lengths $\{l_I\}$ by Heron's formula as in the Regge geometry. Inverting the relation between $\{j_{h_o}\}_{h_o=1}^M$ and $\{l_I\}_{I=1}^M$ defines the local change of variables $(j_{h_o},j_{\bar{h}})\to (l_I,j_{\bar{h}})$ in a neighborhood of the real critical point. This procedure is just changing variables and does not impose any restriction.}) in a neighborhood of $\{\mathring{j}_{h_o}\}$ of a real critical point $\{\mathring{j}_h,\mathring{g}_{ve},\mathring{\bf z}_{vf}\}$, and we have $\rmd^{M+N} j_h=\cj_l\rmd^M l_I\,\rmd^{F-M} j_{\bar{h}}$ where $\cj_l$ is the jacobian. 
\be
A(\Delta)&=&\sum_{\{k_h\}}\int\prod_{I=1}^M \rmd l_I \cz^{\{k_h\}}_\Delta\lt(l_I\rt),\label{integralFormAmp0}\\
\!\!\!\!\!\!\cz_\Delta^{\{k_h\}}\lt(l_I\rt)&=&\int \prod_{\bar{h}}\mathrm{d} j_{\bar h}\, \prod_h\lt(2\l \bm{d}_{\l j_{h}}\rt)\int [\rmd g\rmd \mathbf{z}] e^{\lambda S^{(k)}}\cj_l,\label{cz0000}
\ee
The partial amplitude $\cz_\Delta^{\{k_h\}}$ has the external parameters $r\equiv\{l_I,j_b,\xi_{eb}\}$ including not only the boundary data $j_b,\xi_{eb}$ but also internal segment-lengths $l_I$. Then we apply the stationary phase analysis for the complex action with parameters \cite{10.1007/BFb0074195,Hormander}\footnote{The literature \cite{10.1007/BFb0074195,Hormander} uses the semi-analytic machinery, since they deal with more generic situation with $S$ being smooth instead of being analytic.} to $\cz_\Delta^{\{k_h\}}$: Consider the large-$\l$ integral $\int_K e^{\l S(r,x)}\rmd^N x$ and regard $r$ as the external parameter. $S(r,x)$ is an analytic function of $r\in U\subset \R^k,x\in K\subset \R^N$. $U\times K$ is a neighborhood of $(\mathring{r},\mathring{x})$, where $\mathring{x}$ is a real critical point of $S(\mathring{r},x)$. We denote by $\mathcal{S}(r,z)$, $z=x+i y \in \mathbb{C}^{N}$, the analytic extension of $S(r,x)$ to a complex neighborhood of $\mathring{x}$. The complex critical equation is given by $\partial_{z} \mathcal{S}=0$, which contains $N$ holomorphic equations for $N$ complex variables. The complex critical equation is solved by ${z}=Z(r)$ where $Z(r)$ is analytic in $r$ in the neighborhood $U$. When $r=\mathring{r}$, $Z(\mathring{r})=\mathring{x}$ reduces to the real critical point. When $r$ deviates away from $\mathring{r}$, $Z(r)\in\C^N$ can move away from the real plane $\R^N$, and thus generally $Z(r)$ is called the \emph{complex critical point} (see Figure \ref{pic}(b)). We have the following large-$\l$ asymptotic expansion for the integral
	\be
	\int_K e^{\lambda S(r,x)}  \mathrm{d}^N x &=& \left(\frac{1}{\lambda}\right)^{\frac{N}{2}} \frac{e^{\lambda \mathcal{S}(r,Z(r))}}{\sqrt{\det\left(-\delta^2_{z,z}\mathcal{S}(r,Z(r))/2\pi\right)}} \lt[1+O(1/\l)\rt]%\sum_{s=0}^{\infty}\left(\frac{1}{\lambda}\right)^{s}\left[L_{s} \tilde{u}\right](Z(b))
	\label{asymptotics}
	\ee
where $\cs(r,Z(r))$ and $\delta^2_{z,z}\mathcal{S}(r,Z(r))$ are the action and Hessian at the complex critical point.

The importance of \eqref{asymptotics} is that the integral can receive the dominant contribution from the complex critical point away from the real plane. This fact has been overlooked by the argument of the flatness problem. Moreover, Eq.\eqref{asymptotics} reduces $A(\Delta)$ to the integral
\be
\left(\frac{1}{\lambda}\right)^{\frac{N}{2}}\!\!\!\int\prod_{I=1}^M \rmd l_I \sn_l\,{e^{\lambda \mathcal{S}(r,Z(r))}} \lt[1+O(1/\l)\rt]\label{pathintegral0}
\ee
at each ${k_h}$. $\sn_l\propto\prod_h\lt(4 j_{h}\rt)\cj_l[\det(-\delta^2_{z,z}\cs/2\pi)]^{-1/2}$ at $Z(r)$. Given that $\{l_I\}$ determines the Regge geometry on $\Delta$, Eq.\eqref{pathintegral0} is a path integral of Regge geometries with the complex effective action $\cs$. The path integral sums over curved geometries. In the following, we make the above general analysis concrete by considering the amplitudes on $\Delta=\Delta_3,\sig_{\text{1-5}}$, and we compute numerically the complex critical points, which demonstrate the curved geometries in the spinfoam amplitude.

%\section{Numerical analysis of $A(\Delta_3)$}

\subsection{Applications}

\subsubsection{Asymptotics of $A(\Delta_3)$}
\label{sec:flatness}

The simplicial complex $\Delta_3$ contains three 4-simplices and a single internal triangle $h$. All line segments of $\Delta_3$ are boundary, so $M=0$ in \eqref{integralFormAmp0}. The Regge geometry $\mathbf{g}$ on $\Delta_3$ is completely fixed by the (Regge-like) boundary data $\{j_b,\xi_{eb}\}$ that uniquely corresponds to the boundary segment-lengths. 

Following the above general scheme, $r=\{j_b,\xi_{eb}\}$ is the boundary data. $\mathring{r}=\{\mathring{j}_b,\mathring{\xi}_{eb}\}$ determines the flat geometry, denoted by $\mathbf{g}(\mathring{r})$, with $\delta_h=0$. $\mathring{x}=\{\mathring{j}_h,\mathring{g}_{ve},\mathring{\bf z}_{vf}\}$ is the real critical point associated to $\mathring{r}$, and it endows the orientations $s_v=+1$ to all 4-simplices. $\mathring{r}$, $\mathbf{g}(\mathring{r})$, and $\mathring{x}$ are computed numerically in \cite{Han:2021kll}. The integration domain of $A(\Delta_3)$ is 124 real dimensional. The local coordinates $x\in\R^{124}$ covers the neighborhood of $\mathring{x}$ inside the integration domain. $S(r,x)$ is the spinfoam action and is analytic in the neighborhood of $(\mathring{r},\mathring{x})$. $z\in\C^{124}$ complexifies $x$. $\cs(r,z)$ extends holomorphically $S(r,x)$ to a complex neighborhood of $\mathring{x}$. Here we only complexify $x$ but do not complexify $r$. We focus on $k_h=0$, since the integrals with $k_h\neq 0$ have no real critical point when $r=\mathring{r}$, and are still suppressed even when taking into account complex critical points, as far as $\delta_h$ is not close to $4\pi k_h$. 

We deform the boundary data $r=\mathring{r}+\delta r$ to obtain the curved geometries $\mathbf{g}(r)$ with $\delta_h\neq 0$. In practice, we vary the length ${l}_{26}$ of the line segment connecting the points 2 and 6, while leaving other segment lengths unchanged. A family of (Regge-like) boundary data $r$ parametrized by ${l}_{26}$ can be constructed numerically, and give the family of curved geometries.

At each $r\neq \mathring{r}$, the real critical point is absent. But we are able to find the complex critical point $z=Z(r)$ satisfying $\partial_{z}\cs(r,z)=0$ with the high-precision numerics. The details about the numerical solution and error analysis are given in \cite{Han:2021kll}. We insert $Z(r)$ into $\cs(r,z)$, and we compute numerically the difference between $\cs(r,Z(r))$ and the Regge action $\ci_R$ of the \emph{curved} geometry $\mathbf{g}(r)$:
\be
\delta \ci(r)&=&\cs(r,Z(r))-i\ci_{R}[\mathbf{g}(r)],\label{diff}\\
\text{where}&&\ci_{R}[\mathbf{g}(r)]=\fa_h(r)\delta_h(r)+\sum_b\fa_b(r)\Theta_b(r).
\ee
The areas $\fa_h(r),\fa_b(r)$ and deficit/dihedral angles $\delta_h(r)$,$\Theta_b(r)$ are computed from $\mathbf{g}(r)$. 

%There is no sum over $h$ since the internal triangle is unique in $\Delta_3$.

%$\sum_b\fa_b\Theta_b$ in the Regge action is determined only by the boundary data.

\begin{figure}[h]
	\centering
	\includegraphics[width=0.9\textwidth]{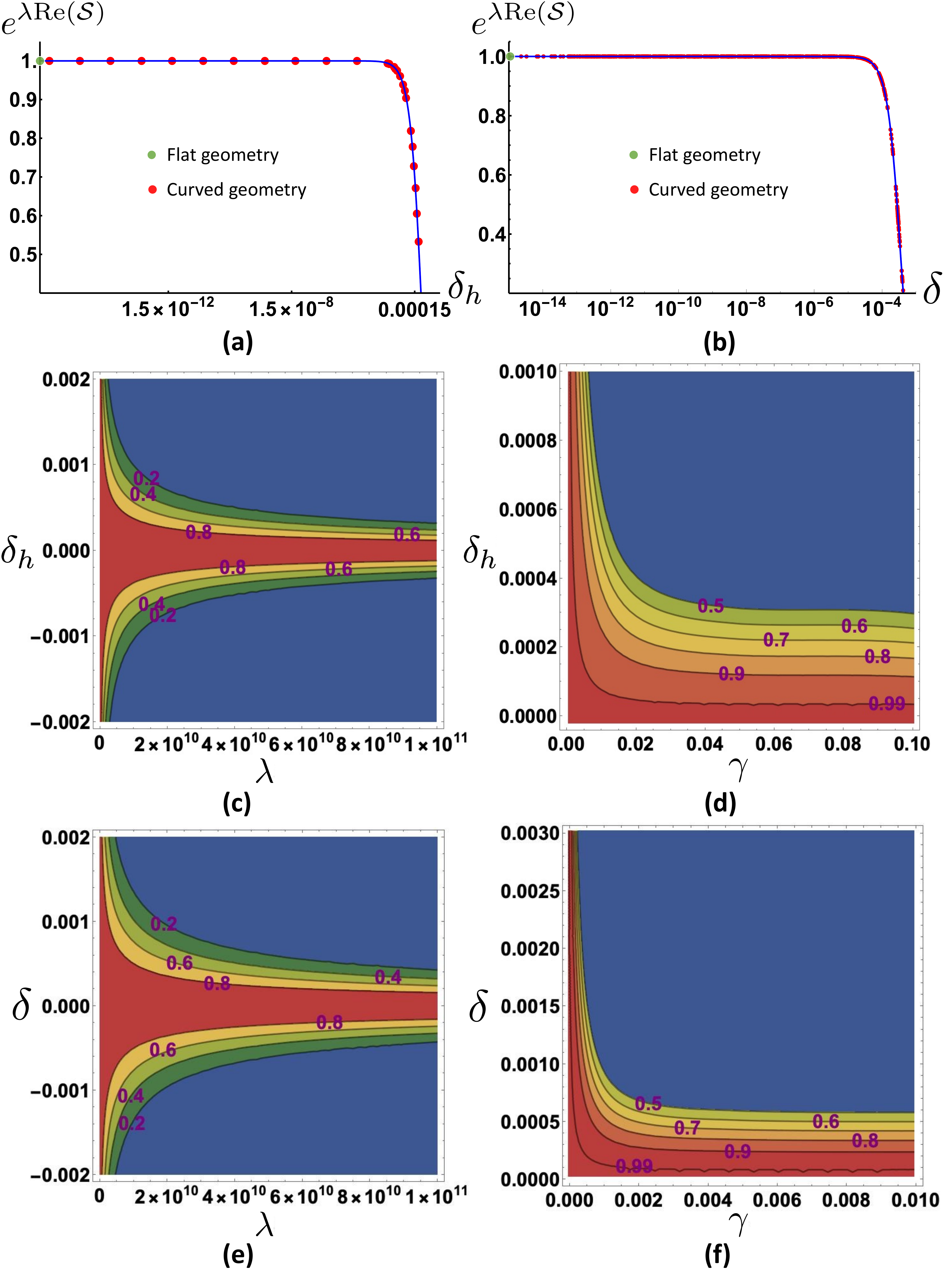}
	\caption{ (a) plots $e^{\lambda\re(\cs)}$ versus the deficit angle $\delta_{h}$ at $\lambda=10^{11}$ and $\gamma=0.1$ in $A(\Delta_3)$, and (b) plots $e^{\lambda\re(\cs)}$ versus the deficit angle $\delta=\sqrt{\frac{1}{10}\sum_{h=1}^{10}\delta_h^2}$ at $\lambda=10^{11}$ and $\gamma=1$ in $\cz_{\sig_{\text 1-5}}$. These 2 plots show the numerical data of curved geometries (red points) and the best fits \eqref{deltaciexp} and \eqref{cspachner} (blue curve). (c) and (d) are the contour plots of $e^{\lambda\re(\cs)}$ as functions of $(\lambda,\delta_h)$ at $\gamma=0.1$ and of $(\g,\delta_{h})$ at $\lambda=5\times10^{10}$ in $A(\Delta_3)$. (e) and (f) are the contour plots of $e^{\lambda\re(\cs)}$ as functions of $(\lambda,\delta)$ at $\gamma=1$ and of $(\g,\delta)$ at $\lambda=5\times10^{10}$ in $\cz_{\sig_{\text{1-5}}}$. They demonstrate the (non-blue) regime of curved geometries where the spinfoam amplitude is not suppressed. 
	}
	\label{thetaS}
\end{figure}

We repeat the computation for many $r$'s from varying the length $l_{26}$. The computations give a family of $\delta\ci(r)$ with many $r$'s. By the asymptotic formula \eqref{asymptotics}, the dominant contribution from $Z(r)$ to $A(\Delta_3)$ is proportional to $|e^{i\l \cs}|=e^{\l \re(\cs)}\leq 1$. As shown in Figure \ref{thetaS}(a) and (c), given any finite $\l\gg1$, there are the curved geometries with small nonzero $|\delta_h|$, such that $|A(\Delta_3)|$ is the same order of magitude as $|A(\Delta_3)|$ at the flat geometry. The range of allowed $\delta_h$ for non-suppressed $A(\Delta_3)$ is nonvanishing as far as $\l$ is finite. The range of allowed $\delta_h$ is enlarged when $\g$ is small, shown in Figure \ref{thetaS}(d). The qualitative behavior is similar to the result from the effective spinfoam model in \cite{Asante:2020qpa}. Moreover, motivated by relating $\delta\ci$ to higher curvature terms \cite{lowE}, we find the best polynomial fit of $\delta \ci$ in terms of $\delta_h$ (the blue curve in Figure \ref{thetaS}(a))
\be
\delta \ci= a_2(\g) \delta_{h}^2+a_3(\g) \delta_{h}^3+a_4 (\g)\delta_{h}^4+ O(\delta_{h}^5),\label{deltaciexp}
\ee
For example, at $\g=0.1$, the best fit coefficient $a_i$ ($i=2,3,4$) and the corresponding fitting errors are $a_2=-0.00016_{\pm10^{-17}}-0.00083_{\pm10^{-16}}i$, $a_3=-0.0071_{\pm10^{-13}}-0.011_{\pm10^{-12}}i$, $a_4=-0.059_{\pm10^{-9}}+0.070_{\pm10^{-8}}i$.

%We remark that the semiclassical behavior of the spinfoam amplitude is given by the $1/\l$ expansion as \eqref{asymptotics} with finite $\l$. It is similar to quantum mechanics, where $\hbarn$ is finite and the classical mechanics is reproduced by the $\hbarn$-expansion. The finite $\l$ always leads to the finite range of nonvanishing $\delta_h$. 

So far we have considered the complex critical point near $\mathring{x}$ with all $s_v=+1$. Given the boundary data $\mathring{r}$, there are exactly 2 real critical points $\mathring{x}$ and $\mathring{x}'$, where $\mathring{x}'$ corresponds to the same flat geometry but with all orientations $s_v=-1$. Other 6 discontinuous orientations (two 4-simplices has plus/minus and the other has minus/plus) do not leads to any real critical point, because they violates the flatness constraint $\g {\delta}^{s}_h=\g\sum_vs_v\Theta_h(v)=0$, and $|{\delta}^{s}_h|$ is not small for the discontinuous orientation, as it can be checked numerically. Their contribution to $A(\Delta_3)$ is suppressed even when considering the complex critical point. Therefore we only need to focus on the integrals over 2 real neighborhoods $K,K'$ of $\mathring{x},\mathring{x}'$, while the integral outside $K\cup K'$ only gives suppressed contribution to $A(\Delta_3)$ for large $\l$. We carry out a similar analysis as the above for the integral over $K'$, and we obtain the following asymptotic formula of $A(\Delta_3)$ with $r=\mathring{r}+\delta r$
\be
A(\Delta_3)&=&\lt(\frac{1}{\l}\rt)^{60}e^{i\varphi}\lt[\sn_+ e^{i\l\ci_R[\mathbf{g}(r)]+\l\delta \ci(r)}+\sn_- e^{-i\l\ci_R[\mathbf{g}(r)]+\l \delta \ci'(r)}\rt]\nonumber\\
&&\lt[1+O(1/\l)\rt].\label{asymp}
\ee
where $\varphi$ is the overall phase. 2 complex critical points near $\mathring{x},\mathring{x}'$ respectively contribute 2 terms, with phase plus or minus the Regge action of $\mathbf{g}(r)$ corrected by $\delta\ci(r)$ and $\delta\ci'(r)=\delta\ci(r)^*|_{\delta_h\to -\delta_h}$. $\sn_\pm$ are proportional to $[\det(-\delta^2_{z,z}\cs/2\pi)]^{-1/2}$ evaluated at these 2 complex critical points.

Known as the cosine problem, there has been the naive guess $A(\Delta_3)\sim (\sn_1 e^{i\l \ci_R}+\sn_2 e^{-i\l \ci_R})^3$ (each factor is from the vertex amplitude), whose expansion gave 8 terms corresponding to all possible orientations (see e.g. \cite{Dona:2020tvv}). But Eq.\eqref{asymp} demonstrates that $A(\Delta_3)$ only contain 2 terms corresponding to the continuous orientations.

\subsubsection{1-5 Pachner move} 

$\sig_{\text{1-5}}$ is the complex of the 1-5 Pachner move for refining one 4-simplex into five 4-simplices. $\sig_{\text{1-5}}$ has $5$ internal segments (see Figure \ref{pic}(a)), so $M=5$ in \eqref{integralFormAmp0}, in contrast to $\Delta_3$, where all segments are on the boundary. There are $10$ internal triangles $h$ in $\sig_{\text{1-5}}$. The integrals in the partial amplitude $\cz_{\sig_{\text{1-5}}}$ have the external parameters $r\equiv\{l_I,j_b,\xi_{eb}\}$ including not only the boundary data $j_b,\xi_{eb}$ but also internal segment-lengths $l_I$ ($I=1,\cdots,5$). We set $\mathring{r}\equiv\{\mathring{l}_I,\mathring{j}_b,\mathring{\xi}_{eb}\}$ determining all internal and boundary segment-lengths of the flat geometry $\mathbf{g}(\mathring{r})$ on $\sig_{\text{1-5}}$. Here we still focus on $k_h=0$, and we find in $\cz_{\sig_{\text{1-5}}}$ the non-degenerate real critical point $\mathring{x}=\{\mathring{j}_{\bar h},\mathring{g}_{ve},\mathring{\bf z}_{vf}\}$ corresponding to the flat geometry with all $s_v=1$.

There are the local coordinates $x\in\R^N$ covering the neighborhood $K$ of $\mathring{x}$. We analytic continue the spinfoam action $S(r,x)$ to $\cs(r,z)$, where $z\in\C^N$ and we fix the boundary data and deform the internal lengths $l_I=\mathring{l}_I+\delta l_I$, so that $r=\{{l}_I,\mathring{j}_b,\mathring{\xi}_{eb}\}\equiv r_l$ give curved geometries $\mathbf{g}(r_l)$. As in \eqref{asymptotics}, the dominant contribution to $A({\sig_{\text{1-5}}})$ from the complex critical point $Z(r_l)$ is given by
\be
\left(\frac{1}{\lambda}\right)^{\frac{155}{2}}\!\!\!\int\prod_{I=1}^5 \rmd l_I \sn_l\,{e^{\lambda \mathcal{S}(r_l,Z(r_l))}} \lt[1+O(1/\l)\rt].\label{pathintegral}
\ee
This formula reduces $A(\sig_{\text{1-5}})$ to the integral over Regge geometries $\mathbf{g}(r_l)$. $\sn_l$ is proportional to $\prod_h\lt(4 j_{h}\rt)\cj_l[\det(-\delta^2_{z,z}\cs/2\pi)]^{-1/2}$ at $Z(r_l)$. The numerical result of $|e^{i\l \cs}|=e^{\l\mathrm{Re}(\cs)}$ is presented in Figure \ref{thetaS} (b), (e), and (f), which demonstrate curved geometries with small $|\delta_h|$ do not lead to the suppression of $\cz_{\sig_{\text{1-5}}}(l_I)$. Moreover $\cs(r_l,Z(r_l))$ in \eqref{pathintegral} is numerically fit by:
\be
\!\!\!\!\!\cs(r_l,Z(r_l))=-i\ci_R[\mathbf{g}(r_l)]-{a_2(\g)}\delta(r_l)^2+O(\delta^3),\label{cspachner}
\ee
where $\delta(r_l)=\sqrt{\frac{1}{10}\sum_{h=1}^{10}\delta_h(r_l)^2}$ and $a_2=8.88\times 10^{-5}_{\pm 10^{-12}} - i 0.033_{\pm 10^{-10}}$ at $\g=1$. $\ci_R[\mathbf{g}(r_l)]$ is the Regge action of $\mathbf{g}(r_l)$. Some more detailed analysis are given in \cite{Han:2021kll}.

%% file: Lefschetz.tex
\subsection{The Lefschetz Thimble and Algorithm}\label{sec:TM}

As shown in Section \ref{sec_integral}, the spinfoam action $S$ is complex valued. As a result, the integrands in \eqref{integralFormAmp1} are highly oscillatory, especially when $\lambda$ is large. This fact plagues the attempts to use the conventional Monte-Carlo method to compute the amplitude and observables in spinfoam. In this section, we review how to use \textit{Picard-Lefschetz theory} to transform these types of integrals with complex actions to non-oscillatory integrals(see e.g. \cite{Witten:2010cx,Alexandru:2020wrj}) for detailed reviews). We summarize the framework presented in \cite{Han:2020npv} that combines the thimble and Markov-Chain Monte Carlo (MCMC) methods which can compute the expectation value of observables when the action is complex valued. 

\subsubsection{Picard-Lefschetz theory and Lefschetz Thimbles}\label{subsec:thimbleframework}
The Lefschetz thimble method is a high-dimensional generalization of the saddle point integration along the steepest descent (SD) paths. 
We start from a multi-dimensional integral, which takes the general form
\be\label{eq:AA}
  A=\int \dd^n x f(\vec{x}) \mathrm{e}^{-S(\vec{x})},
\ee
with $S(\vec{x})$ complex valued. The starting point
is to analytically continue both $f(\vec{x})$ and $S(\vec{x})$ to holomorphic functions $\hat{f}(\vec{z})$ and $\hat{S}(\vec{z})$. Equation \eqref{eq:AA} becomes an integral of analytic functions $\hat{f}(\vec{z})$ and $\hat{S}(\vec{z})$ of complex variables on the real domain
  \be
  A=\int_{\mathbb{R}^n} \dd^n z \hat{f}(\vec{z}) \mathrm{e}^{-\hat{S}(\vec{z})},
  \ee  
where $\dd^n z \hat{f}(\vec{z}) \mathrm{e}^{-\hat{S}(\vec{z})}$ is a holomorphic $n$-form restricted on $\R^n$.
  
The Picard-Lefschetz theory shows that the integral $A$ can be decomposed into a linear combination of integrals over real $n$-dimensional integral cycles $\mathcal{J}_\sigma,\ \sigma=1,2,3,\cdots$ on the complex domain
  \be\label{eq:decomT}
    \int_{\mathbb{R}^n} \dd^n z \hat{f}(\vec{z}) \mathrm{e}^{-\hat{S}(\vec{z})}=\sum_{\sigma} n_\sigma \int_{\mathcal{J}_\sigma} \dd^n z \hat{f}(\vec{z}) \mathrm{e}^{-\hat{S}(\vec{z})},
  \ee
where  $\sum_\sigma n_\sigma \mathcal{J}_\sigma$ with weights $n_\sigma$ is homologically equivalent to $\mathbb{R}^n$. %The holomorphic $n$-form $\dd^n z \hat{f}(\vec{z}) \mathrm{e}^{-\hat{S}(\vec{z})}$ is restricted in a class of ${\mathcal{J}_\sigma}$ on the right hand side.  
The $n$-dimensional real sub-manifolds $\{ \mathcal{J}_{\sigma} \}$, called Lefschetz thimbles, present a good basis of relative homology group for the integral \eqref{eq:AA} \cite{Scorzato:2015qts}. 
Each  $\mathcal{J}_{\sigma}$ is attached to a critical point $p_{\sigma}$ satisfying $\partial_z\hat{S}(p_{\sigma})=0$ in the complex space\footnote{We do not impose $\mathrm{Re}(\hat{S})=0$ for critical points in the complex space.}, as a union of SD paths that are solutions to the SD equations 
  \begin{equation}\label{eq:SDeq}
      \frac{\dd z^a }{ \dd t}=-\frac{\partial\overline{\hat{S}(\vec{z})}}{\partial\overline{z^a}}.
  \end{equation}
Here we call $t$ the flow time. Any points on the thimble will fall to the critical point $p_\sigma$ after flowing for infinitely long time. Following \eqref{eq:SDeq}, $\mathrm{Re}(\hat{S})$ monotonically decreases along each SD path and approaches its minimum at the critical point $p_\sigma$ in the limit $t\to\infty$, while $\mathrm{Im}(\hat{S})$ is a constant along the path.
 Thus, on each thimble $\mathcal{J}_{\sigma}$, the original integral becomes a non-oscillatory integral times a constant phase $\mathrm{e}^{-i \, {\rm Im}{(\hat{S}(p_{\sigma}))}}$,
  \be\label{eq:integralonThimble}
  \int_{\mathcal{J}_\sigma} \dd^n z \hat{f}(\vec{z}) \mathrm{e}^{-\hat{S}(\vec{z})}=\mathrm{e}^{-i \, {\rm Im}{(\hat{S}(p_{\sigma}))}}\int_{\mathcal{J}_\sigma} \dd^n z \hat{f}(\vec{z}) \mathrm{e}^{-{\rm Re}{(\hat{S}(\vec{z}))}} \,.
  \ee
 On each thimble $\mathcal{J}_\sigma$, ${\rm Re}(\hat{S})$ grows when moving far away from the critical point, so the integrand is exponentially suppressed at infinity.
  
 Using the Lefschetz thimble basis $\{ \mathcal{J}_{\sigma} \}$, the integral \eqref{eq:decomT} is valid for a specific set $\{ n_\sigma \}$ of the weights of the thimbles.
   Consider $\hat{f}$ as an observable. The expectation value $\langle f \rangle$ is given by
  \be\label{eq:obs1}
  \langle f \rangle
  =\frac{\int_{\mathbb{R}^n} \dd^n z \hat{f}(\vec{z}) \mathrm{e}^{-\hat{S}(\vec{z})}}{\int_{\mathbb{R}^n} \dd^n z  \mathrm{e}^{-\hat{S}(\vec{z})}}
  =\frac{\sum_{\sigma} n_\sigma \int_{\mathcal{J}_\sigma} \dd^n z \hat{f}(\vec{z}) \mathrm{e}^{-\hat{S}(\vec{z})}}{\sum_{\sigma} n_\sigma \int_{\mathcal{J}_\sigma} \dd^n z  \mathrm{e}^{-\hat{S}(\vec{z})}}.
  \ee
A weight $n_\sigma$ is the intersection number between the original integration cycle $\mathbb{R}^n$ and the manifold of the steepest ascent (SA) paths which are solutions to the SA equations
  \be\label{eq:SA}
    \frac{\dd z^a }{ \dd t}=\frac{\partial\overline{\hat{S}(\vec{z})}}{\partial\overline{z^a}}.
  \ee
Contrary to the SD path, along each SA path $\mathrm{Re}(\hat{S})$ monotonically increases and approaches the maximum at the critical point when $t \to \infty$, with $\mathrm{Im}(\hat{S})$ being conserved. Computing these weights are challenging in general (see e.g. \cite{Bedaque:2017epw,Bluecher:2018sgj} for some recent progresses). Nevertheless, in the case where one of the thimbles $\mathcal{J}_{\sigma'}$ dominates the integral, we may neglect the contribution of other thimbles and re-express \eqref{eq:obs1} as

  \be\label{eq:regThimble}
  \langle f \rangle
  \simeq \frac{n_{\sigma'} \mathrm{e}^{-\mathrm{i}\, {\rm Im}(S(p_{\sigma'}))} \int_{\mathcal{J}_{\sigma'}} \dd^n z\, \hat{f}(\vec{z}) \mathrm{e}^{-{\rm Re} (\hat{S}(\vec{z}))}}{ n_{\sigma'} \mathrm{e}^{-\mathrm{i}\, {\rm Im}(S(p_{\sigma'}))} \int_{\mathcal{J}_{\sigma'}} \dd^n z\,  \mathrm{e}^{-{\rm Re}(\hat{S}(\vec{z}))}}
  =\frac{ \int_{\mathcal{J}_{\sigma'}} \dd^n z\, \hat{f}(\vec{z}) \mathrm{e}^{-{\rm Re} (\hat{S}(\vec{z}))}}{ \int_{\mathcal{J}_{\sigma'}} \dd^n z \, \mathrm{e}^{-{\rm Re}(\hat{S}(\vec{z}))}},
  \ee  
  which can be considered as a mean value under a sampling on the thimble $\mathcal{J}_{\sigma'}$ with a Boltzmann factor $\e^{-{\rm Re}(\hat{S}(\vec{z}))}$. Then, it is possible to use the MCMC method to numerically compute $\langle f \rangle$ in such case.

  %Each integral involved in the spinfoam propagator has a single critical point in its integration domain. The Lefschetz thimble of the critical point is dominant. Thus the computation of the spinfoam propagator is the case where Eq. \eqref{eq:regThimble} applies.
%\end{itemize}

\subsubsection{Thimbles Generated by Flows}\label{ssec:TGF}
%Generating Thimble by morse flow

\begin{figure}[h]
  \centering\includegraphics[width=0.95\textwidth]{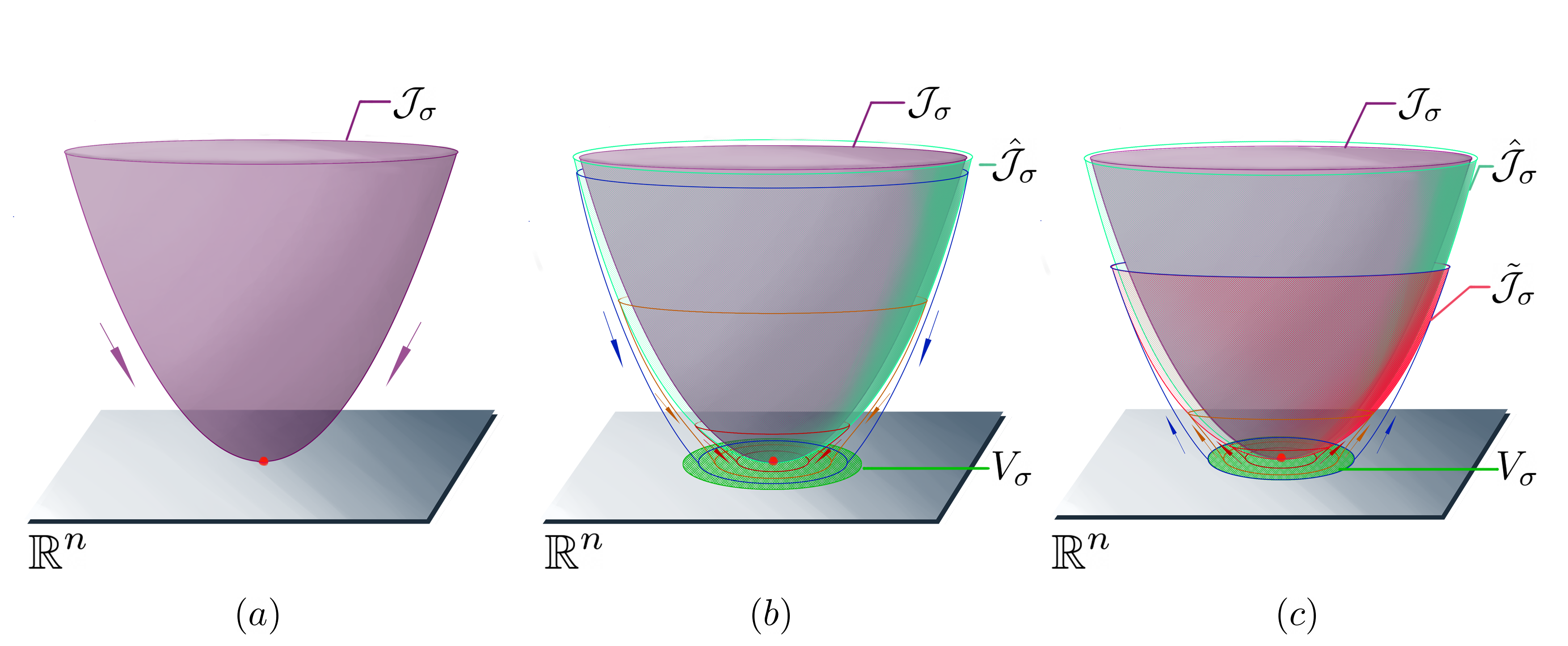}
  \caption{
  (a) A Lefschetz thimble $\mathcal{J}_{\sigma}$ (purple surface) is the union of all the SD paths falling to the critical point $p_\sigma$ (red dot) when $t\to\infty$. \\
  (b) $\widehat{\mathcal{J}}_\sigma$ (green transparent surface) is defined as the union of points that flow to $V_\sigma$ (green disk at the bottom) by the SD equation when $t\to\infty$. The cross-sections of $\widehat{\mathcal{J}}_\sigma$ (illustrated by the blue, yellow and red circles in $\widehat{\mathcal{J}}_\sigma$) flow to the cross-sections in $V_\sigma$ (blue, yellow and red circles in the green disk). \\
  (c) $\tilde{\mathcal{J}}_\sigma$ (red transparent surface) is generated by upward flows from each point in $V_\sigma$ with a finite time by the SA equation. The cross-sections in $V_\sigma$ (blue, yellow and red circles in the green disk) flow upward to the cross-sections in $\tilde{\mathcal{J}}_\sigma$ (blue, yellow and red circles in $\tilde{\mathcal{J}}_\sigma$).
  }\label{fig:thimble1}
\end{figure}

As the first step to applying the above Lefschetz thimble technique, we need to identify the Lefschetz thimble $\mathcal{J}_\sigma$ associated with a given critical point $p_\sigma$ (FIG. \ref{fig:thimble1} (a)). By definition, one might try to decide if a point is on the thimble by checking if it falls to $p_\sigma$ at $t \to \infty$ under the SD equation. However, it is really hard in practice due to the infinite flow time. It is also problematic to use the SA equation with $p_\sigma$ as the initial point to generate the thimble since $p_\sigma$ is a fixed point of the SA equation. 

We follow the method reviewed in \cite{Alexandru:2020wrj} to bypass the difficulty of generating the thimble $\mathcal{J}_{\sigma}$ numerically. The trick is, instead of $p_\sigma$ itself, we consider a small real $n$-dimensional neighborhood $V_{\sigma}$ of $p_\sigma$ and a slightly different integration cycle denoted by $\widehat{\mathcal{J}}_{\sigma}$ (FIG. \ref{fig:thimble1} (b)), which is the union of solutions to the SD equations \eqref{eq:SDeq} falling to $V_{\sigma}$ at $t \to \infty$. %$\widehat{\mathcal{J}}_{\sigma}$ is also real $n$- dimensional. 
$\widehat{\mathcal{J}}_{\sigma}$ is a good approximation of the true thimble $\mathcal{J}_{\sigma}$ when the size of $V_{\sigma}$ is small enough, and it approaches $\mathcal{J}_{\sigma}$ when $V_{\sigma}$ shrinks to the critical point $p_\sigma$. Since the integrand is analytic, and $\widehat{\mathcal{J}}_{\sigma}$ is a deformation of $\mathcal{J}_{\sigma}$, the integral 
\be\label{eq:appo1}
\int_{\widehat{\mathcal{J}}_{\sigma}} d^n z\hat{f}(\vec{z})e^{-\hat{S}(\vec{z})}
\ee
is the same as \eqref{eq:integralonThimble}. However, in this case the above integral becomes oscillatory in contrast to the integral on $\mathcal{J}_{\sigma}$, as ${\rm Im}{(\hat{S})}$ is no longer constant in $\widehat{\mathcal{J}}_{\sigma}$. If $V_{\sigma}$ is small enough, we can control the fluctuation of the ${\rm Im}{(\hat{S})}$ on $\widehat{\mathcal{J}}_{\sigma}$, such that it is still small and the oscillation of the integral is weak enough to keep the Monte Carlo method accurate.

Since infinite time evolution is involved, finding the entire $\widehat{\mathcal{J}}_{\sigma}$ is still not numerically practical. A practical integral cycle $\tilde{\mathcal{J}}_{\sigma}$ is the union of the paths which are solutions to the SD equations \eqref{eq:SDeq} falling to $V_{\sigma}$ at some  finite but sufficiently long flow time $T$. The thimble $\tilde{\mathcal{J}}_{\sigma}$ approaches $\widehat{\mathcal{J}}_{\sigma}$ in the limit $T \to \infty$. Similar to the method in \cite{Bedaque:2017epw}, we can find the approximation $\tilde{\mathcal{J}}_{\sigma}$ as an inverse process using SA flows. We firstly choose a small real $n$-dimensional neighborhood $V_{\sigma}$ of the critical point $p_{\sigma}$, then generate the upward flows from $V_{\sigma}$ according to the SA equation with a finite time $T$. The endpoints of these flows form the real $n$-dimensional manifold $\tilde{\mathcal{J}}_{\sigma}$ (FIG. \ref{fig:thimble1}c). Due to the finite flow time, the thimble $\tilde{\mathcal{J}}_{\sigma}$ does not reach the infinity of the Lefschetz thimble $\mathcal{J}_\sigma$. Its size depends on the choice of flow time $T$. 

As a summary, the approximation of $\mathcal{J}_{\sigma}$ is illustrated in the following diagram: 
\[
\mathcal{J}_\sigma\xrightarrow{\text{Fix } V_{\sigma}}\widehat{\mathcal{J}}_\sigma\xrightarrow{\text{Fix } T}\tilde{\mathcal{J}}_\sigma.
\]
In the first step, we use the $\widehat{\mathcal{J}}_\sigma$ as the union of all the SD paths falling to $V_{\sigma}$ to approximate $\mathcal{J}_{\sigma}$. The size of $V_{\sigma}$ can be set by a tolerance $\mathcal{E}$ of the fluctuation of the ${\rm Im}{(\hat{S})}$ around ${\rm Im}{(\hat{S(p_{\sigma}})}$ on $\widehat{\mathcal{J}}_\sigma$. In the second step, we use $\tilde{\mathcal{J}}_\sigma$ as the union of the finitely evolved SA paths with time $T$ starting from the points in $V_{\sigma}$ to approximate $\widehat{\mathcal{J}}_\sigma$. The longer $T$ and smaller $V_{\sigma}$ are, the better approximation is achieved.

Note that in the second step of the approximation, making $\tilde{\mathcal{J}}_\sigma$ very large (thus a very long $T$) is unnecessary. Recall that when computing \eqref{eq:appo1}, we sample the points on the thimble with the probability distribution $\e^{-{\rm Re}{(S)}}$ where ${\rm Re}{(S)}$ monotonically increases along the SA flow. Thus the contributions from points far away from the critical point are exponentially suppressed. As a result, it is sufficient to choose flow time $T$, which provides $\tilde{\mathcal{J}}_\sigma$ containing points that contribute dominantly to \eqref{eq:appo1}, thus increasing $T$ adds only negligible contributions. The result shall converge when further increasing $T$. In fact, existing results \cite{Alexandru:2020wrj,Alexandru:2015sua,Alexandru:2015xva,Han:2020npv} suggest $T<1$ may already be sufficient for a good accuracy.

%%Tangent space and Takagi vectors

The choice of real $n$-dimensional $V_{\sigma}$ depends on the local behavior of the SA equations \eqref{eq:SA} around the critical point $p_\sigma$.
Consider a small holomorphic variation $\omega^k=\delta z^k$, we can linearize \eqref{eq:SA} around $p_\sigma$:
\be\label{eq:jacobian}
    \frac{\dd \omega^k}{\dd t}=\overline{{\mathbf{H}} \cdot \omega^l},
\ee
where ${\mathbf{H}}={\frac{\partial^2 \hat{S}}{\partial z_k \partial z_l }}|_{z=p_{\sigma}}$ of $\hat{S}(\vec{z})$ is the Hessian matrix at $p_\sigma$. The solution of \eqref{eq:jacobian} is
\[
{\omega}=\sum_{a=1}^{2n}\e^{t \lambda^a} {\omega}_a,
\]
where $\lambda^a$ and ${\omega}_a$ are the eigenvalues and corresponding eigenvectors of the generalized eigenvalue equation: 
\be\label{eq:HES1}
{\mathbf{H}} {\omega} = \overline{ \lambda {\omega}}.
\ee
\eqref{eq:HES1} is a real $2n$-dimensional eigenvalue equation with the Takagi factorization \cite{192483}:
\be\label{eq:HES2}
\left[
\begin{matrix}
  \mathbf{H}_{\mathbb{R}}&&-\mathbf{H}_{\mathbb{I}}\\
-\mathbf{H}_{\mathbb{I}}&&-\mathbf{H}_{\mathbb{R}}\\
\end{matrix}
\right]
\left[
\begin{matrix}
  {\omega}_{\mathbb{R}}\\
 {\omega}_{\mathbb{I}}\\
\end{matrix}
\right]
=\lambda \left[
 \begin{matrix}
 {\omega}_{\mathbb{R}}\\
 {\omega}_{\mathbb{I}}\\
 \end{matrix}
\right],
\ee
where $\mathbf{H}_{\mathbb{R}}$ and $\mathbf{H}_{\mathbb{I}}$ are the real and imaginary parts of the Hessian $\mathbf{H}$. The eigenvalues of this matrix are real and appear in pairs $(-\lambda,\lambda)$. The complex eigenvectors $\{ {\omega}_a \}$ can be reconstructed from the real eigenvectors $({\mathbb{\omega}_a}{}_{\mathbb{R}},{\mathbb{\omega}_a}{}_{\mathbb{I}})$ by
\[
 {\omega}_a={{\omega}_a}{}_{\mathbb{R}}+i  {{\omega}_a}{}_{\mathbb{I}}.
\]
The flow given by \eqref{eq:SA} is repulsive along the eigenvectors $\mathbb{\omega}_a$ with positive eigenvalues, and is attractive with negative eigenvalues. The paths along the attractive directions converge to the critical point $p_\sigma$, so they can not form $\tilde{\mathcal{J}}_\sigma$. Only the paths along the repulsive directions form the $\tilde{\mathcal{J}}_\sigma$. The space $\hat{V}_\sigma$ at $p_\sigma$ in the $\vec{z}$-coordinate chart can be expressed as 
\be\label{eq:basis}
\hat{V}_\sigma=\{ \vec{z}|\vec{z}=\sum_{a=1}^n {\hat{\omega}}_a x^a+p_\sigma,\ \text{each }{x}^a\in\mathbb{R}\text{ is small}\},
\ee 
where ${\hat{\omega}}_a$ is the normalized eigenvectors with positive eigenvalues.

In the active point of view, every point in $\hat{V}_{\sigma}$ flows upward to $\tilde{\mathcal{J}}_\sigma$ according to the SA equation \eqref{eq:SA} with a fixed time $T$. There exists a local diffeomorphism $\mathcal{C}_T:\hat{V}_{\sigma}\to\tilde{\mathcal{J}}_\sigma$ that maps the initial point $p\in\hat{V}_{\sigma}$ to the endpoint $\mathcal{C}_T(p)\in\tilde{\mathcal{J}}_\sigma$. The change from $\{x^i\}$ to $\{z^i\}$ is induced by $\mathcal{C}_T$.
The Jacobian matrix $J^k_i\equiv \partial z^k/\partial x^i$ for the change is determined by the flow equation
\be\label{eq:jacobian1}
    \frac{\dd (J_{i}^{k})_t}{\dd t}=\sum_{l=1}^n\overline{\frac{\partial^2 \hat{S}(\vec{z})}{\partial z_k \partial z_l }} (\overline{J_{i}^{l})_t}.
\ee
%The solution $J_t$ is the Jacobian matrix of a flow of coordinate changes from $\{x^i\}$ to $\{z^i\}$. 
The initial condition $J_0$ is formed by column vectors $\hat{\omega}_a$. Note that $\det(J)$ here is a complex number.

As a result, for any given observable $f$, its expectation value can be computed by 
\be\label{eq:exc}
\begin{split}
\langle f \rangle& 
\simeq \frac{ \int_{\tilde{\mathcal{J}}_{\sigma}} \dd^n z \,\hat{f}(z)\, \mathrm{e}^{- \hat{S}(z)}}{ \int_{\tilde{\mathcal{J}}_{\sigma}} \dd^n z \, \mathrm{e}^{-\hat{S}(z)}}
%&= \frac{ \int_{\hat{V}_{\sigma'}} \dd^n {z'} \e^{\log(\det(J))}\, \hat{f}\, \mathrm{e}^{- \hat{S}}}{ \int_{\hat{V}_{\sigma'}} \dd {z'}^n \e^{\log(\det(J))}  \mathrm{e}^{-\hat{S}}}\\
= \frac{ \int_{\hat{V}_{\sigma}} \dd^n {x}\, \det(J(x))\, \hat{f}\lt(\mathcal{C}_T(x)\rt)\, \mathrm{e}^{- \hat{S}\lt(\mathcal{C}_T(x)\rt)}}{ \int_{\hat{V}_{\sigma}} \dd^n {x}\, \det(J(x))\,  \mathrm{e}^{-\hat{S}\lt(\mathcal{C}_T(x)\rt)}}\\
&= \frac{ 
  \int_{\hat{V}_{\sigma}} \dd^n {x}\, \e^{i (\arg(\det(J))-{\rm Im}(\hat{S}))}\, \hat{f}\, \mathrm{e}^{- {\rm Re}(\hat{S}) + \log(|\det(J)|)} 
}{ 
  \int_{\hat{V}_{\sigma}} \dd^n  {x}\, \e^{i (\arg(\det(J))-{\rm Im}(\hat{S}))}\, \mathrm{e}^{- {\rm Re}(\hat{S}) + \log(|\det(J)|)}
},\\
\end{split}
\ee
For any observable $f$, we define its expectation with respect to a real effective action $S_{eff}$ as 
\be
\langle f \rangle_{eff}=\frac{ \int_{\hat{V}_{\sigma}} \dd^n {x} \,f\, \mathrm{e}^{- S_{eff}} }{ \int_{\hat{V}_{\sigma}} \dd^n {x}\, \mathrm{e}^{- S_{eff}}}.
\ee
\eqref{eq:exc} can be rewritten as
\be\label{eq:effs}
\langle f \rangle \simeq \frac{ 
  \int_{\hat{V}_{\sigma}} \dd^n {x}\, \hat{f}\,\e^{i  \theta_{res}}\,  \mathrm{e}^{- S_{eff}}
}{ 
  \int_{\hat{V}_{\sigma}} \dd^n {x}\,  \mathrm{e}^{-S_{eff}}
} \times\frac{ 
  \int_{\hat{V}_{\sigma}} \dd^n {x}\,  \mathrm{e}^{-S_{eff}} 
}{ 
  \int_{\hat{V}_{\sigma}} \dd^n {x} \,\e^{i  \theta_{res}} \,\mathrm{e}^{- S_{eff}}
}
=\frac{\langle \e^{i \theta_{res}} \hat{f} \rangle_{eff}}{\langle \e^{i \theta_{res}} \rangle_{eff}}.
\ee
with $S_{eff} \equiv {\rm Re}(\hat{S}) - \log(\det(J))$ as the purely real effective action, and $\theta_{res} \equiv \arg(\det(J))-{\rm Im}(\hat{S})$ is the residual phase. Note that the efficacy of MCMC method in computing both $\langle \e^{i \theta_{res}} \hat{f} \rangle_{eff}$ and $\langle \e^{i \theta_{res}} \rangle_{eff}$ relies on the little fluctuating residue phase $\theta_{res}$. The $\arg(\det(J))$ term usually does not have strong fluctuations, e.g. in the spinfoam calculations. The fluctuations coming from ${\rm Im}(\hat{S}(\vec{z}))$ in $\hat{V}_{\sigma}$ can be bounded by a maximal tolerance $\mathcal{E}$. The tolerance determines the size of $\hat{V}_\sigma$, s.t. at any point $p\in\hat{V}_\sigma$, $|\mathrm{Im}(\hat{S}(p))-\mathrm{Im}(\hat{S}(p_\sigma))|\leq \mathcal{E}$.

\subsection{Spinfoam on a Lefschetz Thimble}
In the previous section, we have described the general algorithm of integrals on Lefschetz thimbles. Here, we apply the Lefschetz thimble to spinfoam model. 

%As we mentioned, the spinfoam propagator can be numerically computed by combining the thimble method and the DREAM algorithm. 
     
%Complexified action
First, we complexify the spinfoam variables $j_{ab}, g_a, z_{ab}$ and analytically continue the integrand in \eqref{integralFormAmp22}. As describe in \ref{complex_critical_point}, the analytic continuation  makes $g^\dagger_{ve}$ and ${\bm{z}_{vf}}^{\dagger}$ independent of $g_{ve}$ and ${\bm{z}_{vf}}$. 
The spin variables $j_{h}$ are also complexified, and the integrands are holomorphic in $j_{h}$. The real scaling parameter $\lambda$ is kept real. For spinfoam vertex amplitude, the analytical continuation renders the integrand and the action, denoted by $\tilde{S}$, holomorphic functions of $54$ complex variables. The thimbles are real $54$-dimensional sub-manifolds in the space of complexified spinfoam variables. The detailed discussion of the analytic continuation of the spinfoam integrands is given in \cite{LowE1} (see also \cite{toappear}). 
     
%large spin approximation
As shown in Section \ref{complex_critical_point}, after analytical continuation, $\tilde{S}$ may have more critical points than $S$ does: complex critical points may exist in addition to the real critical points discussed above. The spinfoam integrals admit decompositions as in \eqref{eq:decomT}, where $\{p_{{\sigma}}\}$ contain both real and complex critical points in the complex domain. Complex critical points $p_{\tilde{\sigma}}$ contribute to the integrals if $n_{\tilde \sigma}\neq 0$, namely, there exist SA paths approaching $p_{\tilde{\sigma}}$ from the real variables. Thus, $n_{\tilde \sigma}\neq 0$ for a complex critical point $p_{\tilde{\sigma}}$ implies that $\mathrm{Re}(\tilde{S}(p_{\tilde{\sigma}}))>0$ since $\mathrm{Re}(\tilde{S})=\mathrm{Re}({S})\geq 0$  in the real domain. Strictly positive $\mathrm{Re}(\tilde{S}(p_{\tilde{\sigma}}))$ implies that when the spinfoam integrals are decomposed as in Eq. \eqref{eq:decomT}, the thimbles $\mathcal{J}_{\tilde{\sigma}}$ associated to complex critical points contribute exponentially small at large $\lambda$.

As discussed in Section \ref{complex_critical_point}, at large $\lambda$, there exist a single geometrical critical point $p_{geo}=(k_0,j_0,g_0,z_0)$ dominates the spinfoam integrals \eqref{integralFormAmp22} with geometrical boundary data. Hence, the single Lefschetz thimble $\mathcal{J}_{geo}$ associated to $p_{geo}$ dominates the decomposition \eqref{eq:decomT} of the spinfoam integrals. Therefore, \eqref{eq:regThimble} is applicable to the partition function and expectation values at large $\lambda$. As a result, the expectation value of spinfoam observables is given by
\be\label{O_spinfoam_singlethimble}
\langle O(j,g,z) \rangle&:=&\frac{\langle W|O(j,g,z)|\Psi_0\rangle}{\langle W|\Psi_0\rangle} \\
&\simeq&\frac{ \int_{\mathcal{J}_{geo}}%\limits_{-\epsilon/\l}^{j^{\rm max}+\epsilon/\l} 
\prod_h\mathrm{d} j_{h}\prod_h 2 \l\,\t_{[-\epsilon,\l j^{\rm max}+\epsilon]}(\l j_h)\int [\rmd g\rmd \mathbf{z}]\, O(j,g,z) e^{\l S^{(k_0)}}}{\int_{\mathcal{J}_{geo}}%\limits_{-\epsilon/\l}^{j^{\rm max}+\epsilon/\l} 
\prod_h\mathrm{d} j_{h}\prod_h 2 \l\,\t_{[-\epsilon,\l j^{\rm max}+\epsilon]}(\l j_h)\int [\rmd g\rmd \mathbf{z}]\, e^{\l S^{(k_0)}}} \,, \notag
\ee
for given boundary states $\Psi_0$.
Equations \eqref{O_spinfoam_singlethimble} capture the contributions of the dominant critical point $p_{geo}$ and include all orders of perturbative $1/\lambda$ corrections.
% Namely, when we expand \eqref{O_spinfoam_singlethimble} as $1/\lambda$ power series at the critical point $p_{geo}$, under the stationary phase approximation, the power series are the same as expanding  (\ref{eq:core4f}) - \eqref{eq:core7f} in $1/\lambda$ (see e.g. \cite{Cristoforetti:2012su} for a general argument).
The approximation leading to \eqref{O_spinfoam_singlethimble} neglects the exponentially suppressed contributions at large $\lambda$. These contributions are (1) integrals with $k \neq 0$ in \eqref{integralFormAmp1}, (2) extending some integrals to infinite such as $\int \dd j_{h} $ on the cover space, and (3) the complex critical points and associated Lefschetz thimbles. 

We have shown that each quantity in the spinfoam partition function and observables can be expressed as the power series $\sum_s a_s\lt(\frac{1}{\lambda}\rt)^s$ plus contributions exponentially suppressed (namely suppressed faster than $O(1/\lambda^N)$ for any integer $N$) at large $\lambda$. Eq. \eqref{O_spinfoam_singlethimble} captures the power series containing all the perturbative quantum corrections while neglecting the exponentially suppressed contributions. The exponentially suppressed contributions may be called non-perturbative corrections, as they contain the sub-dominant thimbles associated with the complex critical points generated by the analytical continuation. In this sense, Eq. \eqref{O_spinfoam_singlethimble} captures all perturbative quantum corrections while neglecting non-perturbative corrections.

It is known that in the traditional stationary phase expansion, the computational complexity grows exponentially when computing the coefficient of higher order $O(1/\lambda^s)$ corrections with larger $s$. In this sense, our method with the Lefschetz thimble is a powerful way to compute the spinfoam observables containing perturbative quantum corrections to all orders.
%%per, non-per
%%new spinfoam model on thimble

Besides, similar to the idea in \cite{Witten:2010zr,Cristoforetti:2012su}, we can consider the integral with the Lefschetz thimble as a new definition of the spinfoam amplitude. 
When generalizing to arbitrary simplicial complex ${\Delta}$, we define the spinfoam amplitude on the Lefschetz thimble by
\be
Z_{\mathcal{J}}=\int_{\mathcal{J}}\dd j_h[\dd g\ \dd z]\,e^{-\lambda \hat{S}_{{\Delta}}},\label{ZJ}
\ee
where $\hat{S}_{{\Delta}}$ is the analytic continuation of the spinfoam action on ${\Delta}$ \cite{LowE1}. Here, $\int\dd j_h$ integrates all internal spins, and $\mathcal{J}$ is the Lefschetz thimble associated with a single critical point. % 
Eq. (\ref{ZJ}) has the advantage of focusing on the contribution from a single critical point of all relevant critical points with non-zero weight and excluding the other ones. In particular, when $\mathcal{J}$ is the thimble of the critical point corresponding to the Lorentzian Regge geometry, it excludes contributions from vector geometries and the geometries with flipping orientations (see \cite{Barrett:2009mw,HZ,Kaminski:2017eew,Liu:2018gfc} for the classification of critical points). In addition, Eq. (\ref{ZJ}) is a better formulation from the computational point of view. Applying the Lefschetz thimble to the spinfoam model is also proposed in the context of coupling to cosmological constant \cite{Haggard:2015yda,Haggard:2015kew,hanSUSY}. 

Given the dominant geometric critical point, the spinfoam amplitdue on the Lefschetz thimble has the same perturbative $1/\lambda$ expansion as the usual definition of the spinfoam amplitude and, in particular, has the same semi-classical limit, as shown in the numerical results in Section \ref{Numerical Results}. The small $\lambda$ behavior of (\ref{ZJ}) is different than the usual spinfoam amplitude because non-perturbative corrections are not negligible at small $\lambda$. 

\subsection{Applications: Spinfoam propagator as an example}\label{Numerical Results}
In this section we will show the result of calculating the Lorentzian EPRL spinfoam propagator on a 4-simplex numerically with Lefschetz thimble MCMC method as a concrete example.
From the connected two-point correlation function $G^{abcd}(x,y) = \left\langle q^{ab}(x) q^{cd}(y) \right\rangle - \left\langle q^{ab}(x) \right\rangle \left\langle q^{ab}(x) \right\rangle$,
the spinfoam propagator is defined by 
\be 
G^{abcd}_{mn} = \left\langle E^a_n\cdot E^b_n E^c_m\cdot E^d_m \right\rangle - \left\langle E^a_n\cdot E^b_n\right\rangle\left\langle E^c_m\cdot E^d_m\right\rangle
\ee
where $E^a_n$ is the flux operator through a face $f_{an}$ dual
to the triangle between the tetrahedra $a$ and $n$.
It is obtained by computing the expectation values $\left\langle E^a_n\cdot E^b_n E^c_m\cdot E^d_m \right\rangle$ and $\left\langle E^a_n\cdot E^b_n\right\rangle$. 
As shown in \cite{propagator,propagator1,propagator2,propagator3}, their definition is given by
\begin{align}
  \langle W|\Psi_0\rangle&\simeq  \int_{-\infty}^{\infty} \mathrm{d}^{10} j \int[\mathrm{d} g \,\mathrm{d} \mathbf{z}]  \,\e^{-\lambda S },\label{eq:core4}\\
  \bra{W}E^a_n\cdot E^b_n E^c_m \cdot E^d_m |\Psi_0\rangle&\simeq \int_{-\infty}^{\infty} \mathrm{d}^{10} j \int [\mathrm{d} g \,\mathrm{d} \mathbf{z}] \, \e^{-\lambda S } (A_{an} \cdot A_{bn}) (A_{cm} \cdot A_{dm}),\label{eq:core5}\\
  \bra{W}E^a_n\cdot E^b_n |\Psi_0\rangle&\simeq \int_{-\infty}^{\infty} \mathrm{d}^{10} j \int [\mathrm{d} g \,\mathrm{d} \mathbf{z}]\, \e^{-\lambda S } (A_{an}  \cdot A_{bn} ),\label{eq:core6}\\
  \bra{W} E^c_m \cdot E^d_m |\Psi_0\rangle& \simeq \int_{-\infty}^{\infty} \mathrm{d}^{10} j \int [\mathrm{d} g \,\mathrm{d} \mathbf{z}] \, \e^{-\lambda S}  (A_{cm}  \cdot A_{dm} )\label{eq:core7} \,,
\end{align}
with 
\be
A_{a b}^{i}=\gamma \lambda j_{a b} \frac{\left\langle\sigma^{i} Z_{b a}, \xi_{b a}\right\rangle}{\left\langle Z_{b a}, \xi_{b a}\right\rangle}, A_{b a}^{i}=-\gamma \lambda j_{a b} \frac{\left\langle J \xi_{b a}, \sigma_{i} Z_{b a}\right\rangle}{\left\langle J \xi_{b a}, Z_{b a}\right\rangle} \,.
\ee 
In this example, Differential Evolution Adaptive Metropolis (DREAM) algorithm, which is a multi-chain MCMC method\cite{VRUGT2016273}, is used. The detailed algorithm and the boundary data used for the numerical computation can be found in \cite{Han:2020npv,spinfoam-git,hzcgit}. The code can update $>1000$ samples per thread per hour
on $X64$ processors (tested on systems with AMD EPYC™ 7742 and AMD EPYC™ 7642). Here the {Barbero-Immirzi parameter} is fixed to $\gamma=-0.1$ and different $\lambda$ are choosen in the computation.

%\subsubsection{Expectation Values}

\begin{figure}[h]
  \centering
    \includegraphics[width=0.75\linewidth]{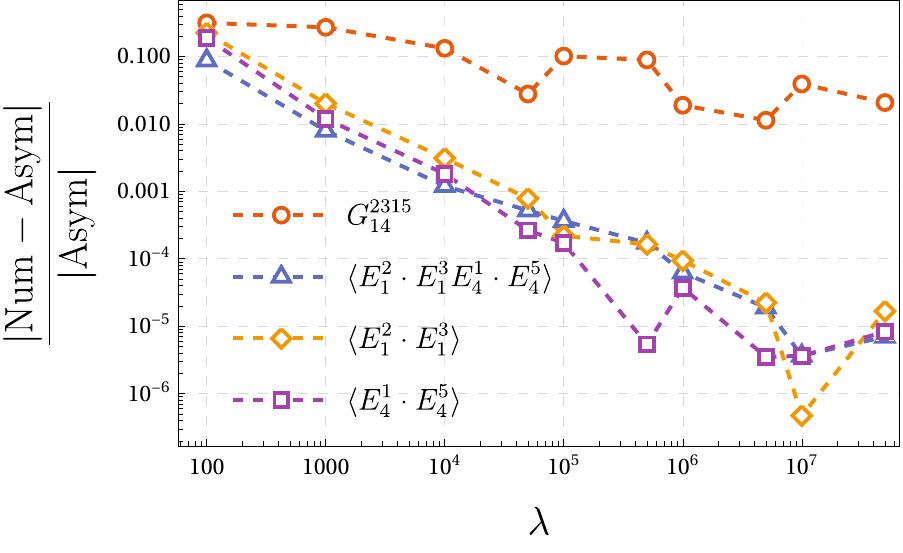}
  \centering
\caption{Difference between the numerical results and the leading order asymptotic results of $G^{2315}_{14}$, $\langle E^2_1\cdot E^3_1 E^1_4 \cdot E^5_4\rangle$, $\langle E^2_1\cdot E^3_1 \rangle$, and $\langle E^1_1\cdot E^5_1 \rangle$}\label{fig:PCA}
\end{figure}

The expectation value of $\left\langle E^a_n\cdot E^b_n E^c_m\cdot E^d_m \right\rangle$ has $1275$ non-zero components and $\left\langle E^a_n\cdot E^b_n\right\rangle$ consists $50$ non-zero components. As a result, the propagator $G^{abcd}_{mn}$ has $1275$ non-zero components. These components are evaluated around $10^7$ of samples obtained by the multi-chain MCMC method on the Lefschetz thimble. As an example of the computation, Figs. \ref{fig:PCA} shows the relative difference between the numerical results and the leading order asymptotic results of the component $G^{2315}_{14}$ and corresponding metric expectation values $\langle E^2_1\cdot E^3_1 E^1_4 \cdot E^5_4\rangle$, $\langle E^2_1\cdot E^3_1\rangle$, and $E^1_4 \cdot E^5_4\rangle$. The results shows the convergence to the leading order asymptotic results in the large spin regime. Their differences become large when $\lambda$ decreases because of the non-negligible contributions from the higher order $1/\lambda$ corrections beyond the leading order asymptotics. 

\begin{figure}[h]
  \centering
  \subfigure[$\lambda=10^6$ with different number of samples]{
  \begin{minipage}[t]{0.5\linewidth}
  \centering
  \includegraphics[width=0.95\linewidth]{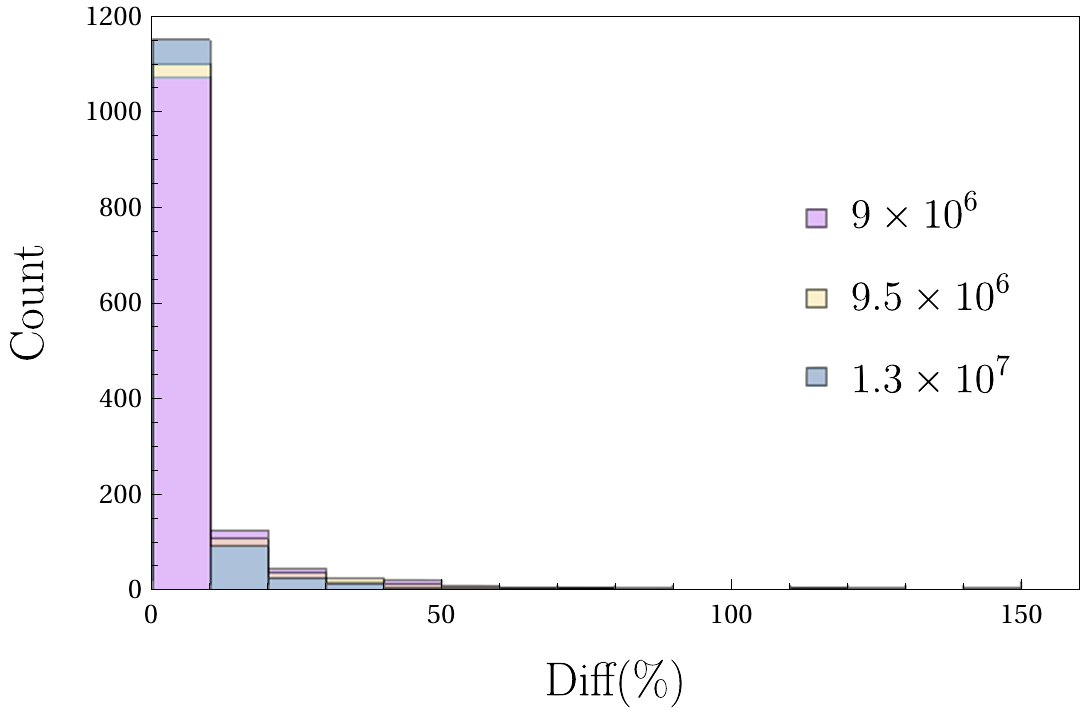}
  %\caption{Abs$\lambda$}\label{fig:EEA}
  \end{minipage}%
  }%
  \subfigure[$\lambda=10^6$ v.s. $\lambda=10^7$]{
  \begin{minipage}[t]{0.5\linewidth}
  \centering
  \includegraphics[width=0.95\linewidth]{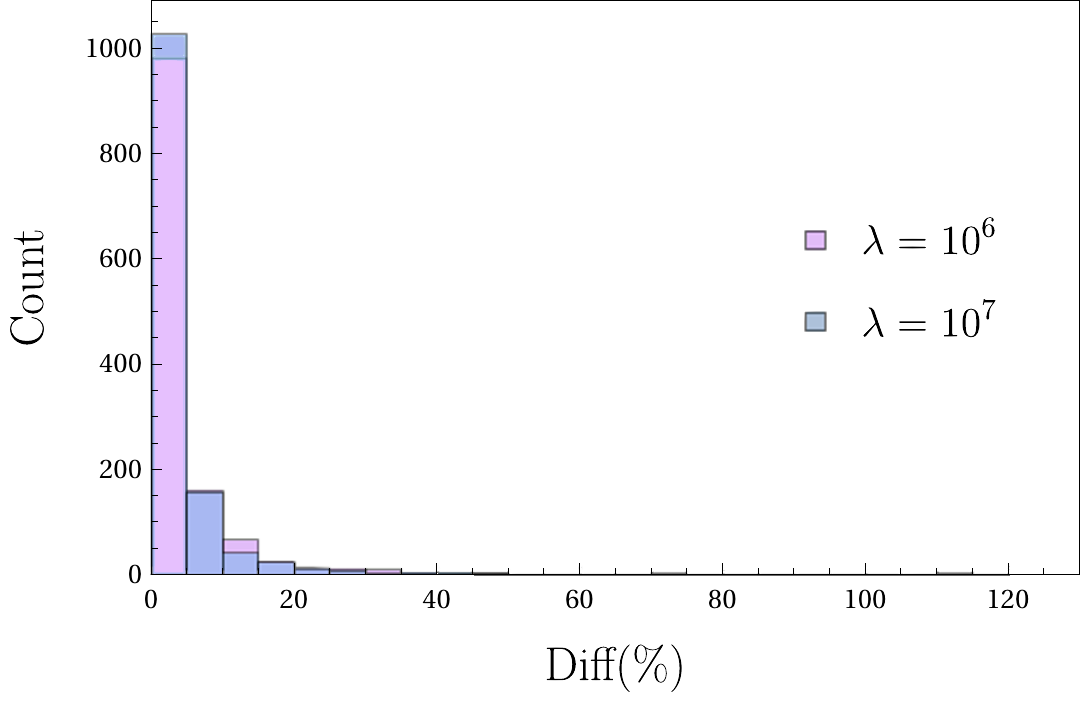}
  %\caption{Arguments $\lambda$}\label{fig:EEB}
  \end{minipage}%
  }%
  \centering
\caption{Histogram of the percentage errors of the components of $G^{abccd}_{mn}$.}\label{fig:PCB}
\end{figure}

The comparison of the results for all of the $1275$ components of the propagator $G^{abcd}_{mn}$ from multi-chain MCMC on the Lefschetz thimble with the results from asymptotic analysis derived in (\ref{asymptotics}) is shown in Figure \ref{fig:PCB}. As shown in Figure \ref{fig:PCB}(a), the percentage differences of these components tend to decrease when the number of samples increase, since increasing the number of samples will make the Markov chains closer to the desired distribution. 
Figure \ref{fig:PCB}(b) compares the histogram of the percentage differences of the components with different $\lambda$ for $\lambda=10^7$ and $\lambda=10^6$. This shows that the Markov chains converge to the desired distribution faster with larger $\lambda$, and the numerical results for larger $\lambda$ are more consistent with the leading order asymptotic results. %(\ref{eq:lead}). 
This fact might suggest that when $\lambda$ becoming large, the less important $1/\lambda$ correction, and the easier converging Markov chains are correlated.

In summary, the algorithm based on Lefschetz thimble MCMC method is able to compute the expectation values of the metric operators as well as the propagators efficiently. The results show the compatibility with the asymptotic result in the large-$\lambda$ regime. As $\lambda$ increasing, the compatibility with the asymptotic results tends to be improved. These results are consistent with the semi-classical behavior of the spinfoam propagator in the large-$\lambda$ regime thus validate the algorithm. When $\lambda$ is not very large, we observe the non-negligible contributions from the higher order $1/\lambda$ corrections beyond the leading order asymptotics results.

%% file: conclusions.tex
Performing calculations has always been arduous in spin foam theory due to its sheer complexity. In recent years the field has undergone a proper numerical revolution. Pulled by the fast developments in high-performance computing, the community developed many different codes and libraries to put a spinfoam in a computer. This chapter reviewed the major frameworks to perform calculations with the Lorentzian EPRL spinfoam model. 

We reviewed first \texttt{sl2cfoam-next} a tool to calculate spin foam amplitudes based on the booster decomposition of the EPRL vertex amplitude. 

The framework has many strengths. It provides all the tools to perform a vast amount of calculations. It has a fast and optimized structure to compute all the necessary building blocks (vertex, face, edge amplitudes, and coherent states). The user needs to compose writing minimal scripts. 
The framework is optimized to run calculations in parallel. Even if we can do small calculations on a personal laptop, most applications we presented require a cluster. The architecture of the library support also tensors contractions using GPUs.  
There are two main problems in the framework. First is the weak control over the truncation parameter, which is necessary to extract numbers from the amplitude with this technique. We do not have a way to prescribe a truncation to match a predetermined and desired error on the amplitude. We are limited to choosing it motivated by empirical arguments. Second is the large number of computational resources needed for extended transition amplitudes. 
These two aspects are currently being studied and improved. We are carrying out a systematical study of possible approximation techniques that would help us to go beyond truncations. Furthermore, we are experimenting with Monte Carlo techniques to overcome the impossible obstacle of the massive amount of resources necessary to sum over bulk degrees of freedom. 
We showcased the power of \texttt{sl2cfoam-next} with some applications. We explored the large spin regime of spinfoam amplitudes with a single and many vertices. We studied the divergences of the theory, and we are currently applying numerical techniques to physical applications like early-time cosmology or black hole tunneling.

Then we reviewed the numerical analysis of the curved Regge geometries from the 4-dimensional Lorentzian EPRL spinfoam amplitude.

The results resolve the flatness problem by explicitly finding the curved Regge geometries emergent from the EPRL spinfoam amplitudes. The curved geometries correspond to the complex critical points away from the real integration domain. They give non-suppressed $e^{\l \re(\cs)}$ and satisfy the bound ${\mathrm{Re}(a_2(\g))} \delta_h^2\lesssim1/\l$, if we neglect high order terms $O(\delta_h^3)$ in dihedral angle $\delta_h$, in the examples we showed. This bound is consistent with the earlier proposal \cite{Han:2013hna} and the result in the effective spinfoam model \cite{Asante:2020qpa,Asante:2020iwm,Asante:2021zzh}, although now this bound should be corrected when taking into account $O(\delta_h^3)$ corrections.% in \eqref{deltaciexp} and \eqref{cspachner}. 
All resulting curved geometries have small deficit angles. The large-$j$ spinfoam amplitude is still suppressed for geometries with larger $\delta_h$. This is not a problem for the semiclassical analysis. Indeed, the non-singular classical spacetime geometries are smooth with vanishing $\delta_h$. To well-approximating smooth geometries by Regge geometries, the triangulation must be sufficiently refined, and all deficit angles must be small.
We showed the example with the method applying to triangulations of the 3-3 ($\Delta_3$) and 1-5 Pachner move. The 1-5 move is one of the elementary moves for the triangulation refinement. Our results provide a new method for analyzing the triangulation dependence in the spinfoam model (see, e.g., \cite{Banburski:2014cwa} for an earlier attempt). This may relate to the spinfoam renormalization \cite{Bahr:2016hwc,Delcamp:2016dqo}, to address the issue of triangulation-dependence of the spinfoam theory. 

Lastly, we reviewed the numerical method for computing the expectation value of any observable based on the Lefschetz thimble and MCMC methods. 

The method applies to all types of spin foam models and graphs, independently of the choice of model (spacetime signature, $Y$-map, etc.). The framework efficiently numerically computes observables in spinfoam models with relatively large spins. It captures perturbative quantum corrections to all orders. The multiple-chain MCMC method used in the framework runs in parallel and supports GPU boost.
One of the main problems in the framework is the generalization to the small spins regime, as we neglect non-perturbative corrections that are exponentially small for large spins in the calculation. Unlike in the asymptotic regime, in the small spin regime, multiple thimbles will contribute to the partition function instead of only one dominant. This requires us to identify all thimbles associated with complex critical points and calculate their weight. Identifying all complex critical points is hard, even in the single simplex case, as the critical equations form a high-order polynomial system with 54 variables. A possible solution to this problem is the world-volume and globally adaptive sampling methods \cite{fukuma2020worldvolume}. We are currently exploring the possibility of using these methods, and some preliminary results are available \cite{Huang:2022plb}.
We showcased the power and efficiency of this framework with the application in calculating the spinfoam propagator as an example. Other exciting applications contain, e.g., the evaluation of geometrical observables in cosmology and black holes, exploring the continuum limit and the renormalization group flow.
In the meantime, we are currently working on improving the framework by choosing a more suitable ODE solver and by applying different Markov chain Monte Carlo techniques, like Hamiltonian Monte Carlo, to sample the amplitudes efficiently.